\pdfoutput=1

\documentclass[twocolumn,epjc3]{svjour3}

\RequirePackage[T1]{fontenc}

\smartqed

\RequirePackage{graphicx}
\RequirePackage{amssymb}
\RequirePackage{epsf}
\RequirePackage{psfig}
\RequirePackage{mathptmx}
\RequirePackage{flushend}
\RequirePackage[numbers]{natbib}
\RequirePackage{hyperref}

\journalname{Nonlinear Dynamics}

\begin{document}

\title{Escapes in Hamiltonian systems with multiple exit channels: Part I}

\author{Euaggelos E. Zotos}

\institute{Department of Physics, School of Science, \\
Aristotle University of Thessaloniki, \\
GR-541 24, Thessaloniki, Greece \\
Corresponding author's email: {evzotos@physics.auth.gr}
}

\date{Received: 12 January 2014 / Accepted: 6 June 2014 / Published online: 27 June 2014}

\titlerunning{Escapes in Hamiltonian systems with multiple exit channels: Part I}

\authorrunning{E. E. Zotos}

\maketitle

\begin{abstract}

The aim of this work is to review and also explore even further the escape properties of orbits in a dynamical system of a two-dimensional perturbed harmonic oscillator, which is a characteristic example of open Hamiltonian systems. In particular, we conduct a thorough numerical investigation distinguishing between trapped (ordered and chaotic) and escaping orbits, considering only unbounded motion for several energy levels. It is of particular interest, to locate the basins of escape towards the different escape channels and connect them with the corresponding escape periods of the orbits. We split our examination into three different cases depending on the function of the perturbation term which determines the number of escape channels on the physical space. In every case, we computed extensive samples of orbits in both the physical and the phase space by integrating numerically the equations of motion as well as the variational equations. In an attempt to determine the regular or chaotic nature of trapped motion, we applied the SALI method as a chaos detector. It was found, that in all studied cases regions of trapped orbits coexist with several basins of escape. It was also observed, that for energy levels very close to the escape value the escape times of orbits are large, while for values of energy much higher than the escape energy the vast majority of orbits escape very quickly or even immediately to infinity. The larger escape periods have been measured for orbits with initial conditions in the boundaries of the escape basins and also in the vicinity of the fractal structure. Most of the current outcomes have been compared with previous related work. We hope that our results will be useful for a further understanding of the escape mechanism of orbits in open Hamiltonian systems with two degrees of freedom.


\end{abstract}

\section{Introduction}
\label{intro}

Escaping particles from dynamical systems is a subject to which has been devoted many studies over the years. Especially the issue of escapes in Hamiltonian systems is directly related to the problem of chaotic scattering which has been an active field of research over the last decades and it still remains open (e.g., [\citealp{BTS96} -- \citealp{BGOB88}, \citealp{CPR75}, \citealp{C90}, \citealp{CK92}, \citealp{E88}, \citealp{JS88}, \citealp{ML02} -- \citealp{PH86}, \citealp{SASL06} -- \citealp{SS10}]). It is well known, that particular types of Hamiltonian systems have a finite energy of escape and for lower values the equipotential surfaces of the systems are close and therefore escape is impossible. For energy levels beyond the escape energy however, these surfaces open creating exit channels through which the particles can escape to infinity. The literature is replete with studies of such ``open" Hamiltonian systems (e.g., [\citealp{BBS09}, \citealp{CKK93}, \citealp{KSCD99}, \citealp{NH01}, \citealp{STN02}, \citealp{SCK95} -- \citealp{SKCD96}]).

Usually, the infinity acts as an attractor for an escape particle, which may escape through different channels (exits) on the equipotential curve or on the equipotential surface depending whether the dynamical system has two or three degrees of freedom, respectively. Therefore, it is quite possible to obtain basins of escape, similar to basins of attraction in dissipative systems or even the Newton-Raphson fractal structures. Basins of escape have been studied in several papers (e.g., [\citealp{BGOB88}, \citealp{C02}, \citealp{KY91}, \citealp{PCOG96}]). The reader can find more details regarding basins of escape in [\citealp{C02}]. The key idea of studying escape of orbits in open dynamical systems is the existence of a chaotic invariant set of orbits embedded in the system and its stable and unstable manifold, where the unstable manifold in fact coincides with the fractal boundary.

One of the most characteristic models for time-independent Hamiltonian systems of two degrees of freedom is undoubtedly the well-known H\'{e}non-Heiles system [\citealp{HH64}]. A huge load of research on the escape properties of this system has been conducted over the years (e.g., [\citealp{AVS01} -- \citealp{AVS03}, \citealp{BBS08}, \citealp{BSBS12}, \citealp{dML99}, \citealp{S07}]). At this point we should emphasize, that all the above-mentioned references on escapes in the H\'{e}non-Heiles system are exemplary rather than exhaustive, taking into account that a vast quantity of related literature exists.

During the last half century, dynamical systems made up of perturbed harmonic oscillators have been extensively used in order to describe local motion (i.e., near an equilibrium point) (e.g., [\citealp{AEFR06}, \citealp{C93}, \citealp{CK98} -- \citealp{CZ12}, \citealp{FLP98a} -- \citealp{HH64}, \citealp{SI79}, \citealp{Z12a}, \citealp{Z12c} -- \citealp{Z14}]). In an attempt to reveal and understand the nature of orbits in these systems, scientists have used either numerical (e.g., [\citealp{CZ12}, \citealp{KV08}, \citealp{ZC12}]) or analytical methods (e.g., [\citealp{CB82}, \citealp{D91}, \citealp{DE91}, \citealp{E00}, \citealp{ED99}]). Furthermore, potentials made up of harmonic oscillators are frequently used in galactic Astronomy, as a first step for distinguishing between ordered and chaotic local motion in galaxies, since it is widely accepted that the motion of stars near the central region of a galaxy can be approximated by harmonic oscillations.

A simple dynamical system of two coupled harmonic oscillators for various values of the energy above the escape energy has been investigated in [\citealp{CE04}], where it was found that stable periodic orbits are surrounded by stability islands that never escape. A further numerical analysis of the same dynamical system in [\citealp{CHLG12}], revealed that as the energy increases beyond the escape value, the majority of chaotic orbits escape either directly, or after a small or large number of intersections with the $y = 0$ axis. In the same vein, the effects of different types of perturbations on both the topology and the escaping dynamics in the H\'{e}non-Heiles system was examined in [\citealp{BSBS12}], where basins of escape were found to exist in the physical $(x,y)$ as well as in the phase $(y,\dot{y})$ space.

Escaping and trapped orbits in stellar systems is an issue of paramount importance. In a recent article [\citealp{Z12b}], we explored the nature of the orbits of stars in a galactic-type potential, which can be considered to describe local motion in the meridional plane $(R,z)$ near the central parts of an axially symmetric galaxy. It was observed, that apart from the trapped orbits there are two types of escaping orbits, those which escape fast and those which need to spend vast time intervals inside the limiting curve before they find the exit and eventually escape. Furthermore, the chaotic dynamics within a star cluster embedded in the tidal field of a galaxy was explored in [\citealp{EJSP08}]. In particular, by scanning thoroughly the phase space and obtaining the basins of escape with the respective escape times it was revealed, that the higher escape times correspond to initial conditions of orbits near the fractal basin boundaries.

Thus, taking into account all the above-mentioned facts, we decided to use a potential of a perturbed harmonic oscillator with such perturbing terms producing between two and four escape channels in the physical $(x,y)$ space. Here we must point out, that these dynamical systems have been studied thoroughly in many previous papers so, in the current work, we shall try to review the main properties of them and also present some more detailed results regarding the escape mechanism of orbits. The aim of this work, is twofold: (i) to distinguish between trapped and escaping orbits and (ii) to locate the basins of escape leading to different escape channels and try to connect them with the corresponding escape times of the orbits. In the forthcoming Part II, we will consider open Hamiltonian systems with $n$ $(n \geq 5)$ channels of escape which however, have not been explored yet.

The present article is organized as follows: in Section \ref{modpot} we describe the properties of the potential we chose for our investigation of trapped and escaping orbits. The computational methods used in order to determine the nature (ordered/chaotic and trapped/escaping) of orbits are described in Section \ref{cometh}. In the following Section, we conduct a thorough analysis of several sets of initial conditions of orbits presenting in detail all the numerical results of our computations. Our article ends with Section \ref{disc}, where the conclusions and the discussion of this research are presented.

\section{Properties of the model potential}
\label{modpot}

The general form of a two-dimensional perturbed harmonic oscillator is
\begin{equation}
V(x,y) = \frac{1}{2}\left(\omega_1^2 x^2 + \omega_2^2 y^2 \right) + \varepsilon V_1(x,y),
\label{genform}
\end{equation}
where $\omega_1$ and $\omega_2$ are the unperturbed frequencies of oscillations along the $x$ and $y$ axes respectively, $\varepsilon$ is the perturbation parameter, while $V_1$ is the function containing the perturbing terms. This is called a two-dimensional perturbed elliptic oscillator.

In the present paper, we shall use a two-dimensional perturbed harmonic oscillator at the 1:1 resonance, that is when $\omega_1 = \omega_2 = \omega$, in order to investigate the escape properties of orbits. The corresponding potential is
\begin{equation}
V(x,y) = \frac{\omega^2}{2}\left(x^2 + y^2 \right) + \varepsilon V_1(x,y),
\label{pot}
\end{equation}
being $\omega$ the common frequency of oscillations along the two axes. Without the loss of generality, we may set $\omega = 1$ and $\varepsilon = 1$ for more convenient numerical computations.

The basic equations of motion for a test particle with a unit mass are
\begin{equation}
\ddot{x} = - \frac{\partial V}{\partial x}, \ \ \
\ddot{y} = - \frac{\partial V}{\partial y},
\label{eqmot}
\end{equation}
where, as usual, the dot indicates derivative with respect to the time. Furthermore, the variational equations governing the evolution of a deviation vector\footnote{If \textbf{$S$} is the $2N$ dimensional phase space where the orbits of a dynamical system evolve on, a deviation vector \vec{w}, which describes a small perturbation of a specific orbit \vec{x}, evolves on a $2N$ dimensional space \textbf{$T_{x}S$} tangent to \textbf{$S$}.} $\vec{w} = (\delta x, \delta y, \delta \dot{x}, \delta \dot{y})$ are
\begin{eqnarray}
\dot{(\delta x)} &=& \delta \dot{x}, \ \ \ \dot{(\delta y)} = \delta \dot{y}, \nonumber \\
(\dot{\delta \dot{x}}) &=& -\frac{\partial^2 V}{\partial x^2}\delta x - \frac{\partial^2 V}{\partial x \partial y}\delta y, \nonumber \\
(\dot{\delta \dot{y}}) &=& -\frac{\partial^2 V}{\partial y \partial x}\delta x - \frac{\partial^2 V}{\partial y^2}\delta y.
\label{variac}
\end{eqnarray}

The Hamiltonian to potential (\ref{pot}) (with $\omega = \varepsilon = 1$) reads
\begin{equation}
H = \frac{1}{2}\left(\dot{x}^2 + \dot{y}^2 + x^2 + y^2\right) + V_1(x,y) = h,
\label{ham}
\end{equation}
where $\dot{x}$ and $\dot{y}$ are the momenta per unit mass conjugate to $x$ and $y$ respectively, while $h > 0$ is the numerical value of the Hamiltonian, which is conserved. The Hamiltonian can also be written in the form
\begin{equation}
H = H_0 + H_1,
\label{ham2}
\end{equation}
with $H_0$ being the integrable term and $H_1$ the non-integrable correction.

Potential (\ref{pot}) has a finite energy of escape $(h_{esc})$ which can be derived as follows: First we solve the system
\begin{equation}
\frac{\partial V}{\partial x} = 0, \ \ \ \frac{\partial V}{\partial y} = 0.
\label{seq}
\end{equation}
The solutions of system (\ref{seq}) gives all the critical points of potential function. The saddle points of (\ref{pot}) are those of the critical points that satisfy the condition
\begin{equation}
S = \left(\frac{\partial^2 V}{\partial x^2}\right) \left(\frac{\partial^2 V}{\partial y^2}\right) - \left(\frac{\partial^2 V}{\partial x \partial y}\right) < 0.
\label{sadl}
\end{equation}
The value of the escape energy is obtained, if we insert the solution of system (\ref{seq}) which satisfy the condition (\ref{sadl}) in the potential (\ref{pot}). It becomes evident, that the escape energy strongly depends on the particular function of the perturbation term $V_1(x,y)$. Here we should note, that in the case when more than one solutions satisfy simultaneously the condition (\ref{sadl}), then $h_{esc}$ is the minimum of the corresponding values of $h$ that are calculated.

\section{Computational methods}
\label{cometh}

In order to study the escape process in our Hamiltonian system, we need to define samples of orbits whose properties (escaping or trapped) will be identified. The best method for this purpose, would have been to choose the sets of initial conditions of the orbits from a distribution function of the system. This however, is not available so, we define for each set of values of the energy (all tested energy levels are above the escape energy), dense grids of initial conditions regularly distributed in the area allowed by the value of the energy. Our investigation takes place both in the physical $(x,y)$ and the phase $(x,\dot{x})$ space for a better understanding of the escape mechanism. In both cases, the step separation of the initial conditions along the $x$ and $y$ and $x$ and $\dot{x}$ axes (in other words the density of the grid) was controlled in such a way that always there are about 50000 orbits (maximum a grid of 225 $\times$ 225 equally spaced initial conditions of orbits).

For each initial condition, we integrated the equations of motion (\ref{eqmot}) as well as the variational equations (\ref{variac}) using a double precision Bulirsch-Stoer FORTRAN algorithm (e.g., [\citealp{PTVF92}]) with a small time step of order of $10^{-2}$, which is sufficient enough for the desired accuracy of our computations (i.e. our results practically do not change by halving the time step). Our previous experience suggests that the Bulirsch-Stoer integrator is both faster and more accurate than a double precision Runge-Kutta-Fehlberg algorithm of order 7 with Cash-Karp coefficients. In all cases, the energy integral (Eq. (\ref{ham})) was conserved better than one part in $10^{-10}$, although for most orbits it was better than one part in $10^{-11}$.

An issue of paramount importance is the determination of the position as well as the time at which an orbit escapes. When the value of the energy $h$ is smaller than the escape energy, the Zero Velocity Curves (ZVCs) are closed. On the other hand, when $h > h_{esc}$ the equipotential curves are open and extend to infinity. An open ZVC consists of several branches forming channels through which an orbit can escape to infinity. At every opening there is a highly unstable periodic orbit close to the line of maximum potential [\citealp{C79}] which is called a Lyapunov orbit. Such an orbit reaches the ZVC, on both sides of the opening and returns along the same path thus, connecting two opposite branches of the ZVC. Lyapunov orbits are very important for the escapes from the system, since if an orbit intersects any one of these orbits with velocity pointing outwards moves always outwards and eventually escapes from the system without any further intersections with the surface of section (see e.g., [\citealp{C90}]). The passage of orbits through Lyapunov orbits and their subsequent escape to infinity is the most conspicuous aspect of the transport, but crucial features of the bulk flow, especially at late times, appear to be controlled by diffusion through cantori, which can trap orbits far vary long time periods.

In our computations, we set $10^5$ time units as a maximum time of numerical integration. Our previous experience in this subject indicates, that usually orbits need considerable less time to find one of the exits in the limiting curve and eventually escape from the system (obviously, the numerical integration is effectively ended when an orbit passes through one of the escape channels and intersects one of the unstable Lyapunov orbits). Nevertheless, we decided to use such a vast integration time just to be sure that all orbits have enough time in order to escape. Remember, that there are the so called ``sticky orbits" which behave as regular ones and their true chaotic character is revealed only after long time intervals of numerical integration. Here we should clarify, that orbits which do not escape after a numerical integration of $10^5$ time units are considered as non-escaping or trapped.

The physical and the phase space are divided into the escaping and non-escaping (trapped) space. Usually, the vast majority of the trapped space is occupied by initial conditions of regular orbits forming stability islands where a third integral is present. In many systems however, trapped chaotic orbits have also been observed. Therefore, we decided to distinguish between regular and chaotic trapped orbits. Over the years, several chaos indicators have been developed in order to determine the character of orbits. In our case, we chose to use the Smaller ALingment Index (SALI) method. The SALI [\citealp{S01}] has been proved a very fast, reliable and effective tool, which is defined as
\begin{equation}
\rm SALI(t) \equiv min(d_-, d_+),
\label{sali}
\end{equation}
where $d_- \equiv \| {\bf{w_1}}(t) - {\bf{w_2}}(t) \|$ and $d_+ \equiv \| {\bf{w_1}}(t) + {\bf{w_2}}(t) \|$ are the alignments indices, while ${\bf{w_1}}(t)$ and ${\bf{w_2}}(t)$, are two deviations vectors which initially point in two random directions. For distinguishing between ordered and chaotic motion, all we have to do is to compute the SALI along time interval $t_{max}$ of numerical integration. In particular, we track simultaneously the time-evolution of the main orbit itself as well as the two deviation vectors ${\bf{w_1}}(t)$ and ${\bf{w_2}}(t)$ in order to compute the SALI. The variational equations (\ref{variac}), as usual, are used for the evolution and computation of the deviation vectors.

The time-evolution of SALI strongly depends on the nature of the computed orbit since when the orbit is regular the SALI exhibits small fluctuations around non zero values, while on the other hand, in the case of chaotic orbits the SALI after a small transient period it tends exponentially to zero approaching the limit of the accuracy of the computer $(10^{-16})$. Therefore, the particular time-evolution of the SALI allow us to distinguish fast and safely between regular and chaotic motion (e.g., [\citealp{ZC13}]). Nevertheless, we have to define a specific numerical threshold value for determining the transition from regularity to chaos. After conducting extensive numerical experiments, integrating many sets of orbits, we conclude that a safe threshold value for the SALI is the value $10^{-7}$. In order to decide whether an orbit is regular or chaotic, one may use the usual method according to which we check after a certain and predefined time interval of numerical integration, if the value of SALI has become less than the established threshold value. Therefore, if SALI $\leq 10^{-7}$ the orbit is chaotic, while if SALI $ > 10^{-7}$ the orbit is regular. For the computation of SALI we used the LP-VI code [\citealp{CMD14}], a fully operational code which efficiently computes a suite of many chaos indicators for dynamical systems in any number of dimensions.

\section{Numerical results}
\label{numres}

Our main objective is to distinguish between trapped and escaping orbits for values of energy larger than the escape energy where the Zero Velocity Curves are open and several channels of escape are present. Moreover, two additional properties of the orbits will be examined: (i) the directions or channels through which the particles escape and (ii) the time-scale of the escapes (we shall also use the term escape period). In the present paper, we explore these aspects for various values of the energy $h$, as well as for three different types of perturbation. The function of the perturbation term plays a key role as it determines the location as well as the number of the escape channels both in the physical and the phase space. In particular, three different cases of perturbation are considered which produces two, three and four channels of escape at the physical $(x,y)$ space respectively. In both cases, the grids of initial conditions of orbits whose properties will be examined are defined as follows: For the physical $(x,y)$ space we consider orbits with initial conditions $(x_0, y_0)$ with $\dot{x_0} = 0$, while the initial value of $\dot{y_0}$ is always obtained from the energy integral (\ref{ham}) as $\dot{y_0} = \dot{y}(x_0,\dot{x_0},h) > 0$. Similarly, for the phase $(x,\dot{x})$ space we consider orbits with initial conditions $(x_0, \dot{x_0})$ with $y_0 = 0$, while again the initial value of $\dot{y_0}$ is obtained from the energy integral (\ref{ham}).

\subsection{Case I: Two channels of escape}
\label{case1}

\begin{figure*}[!tH]
\centering
\resizebox{\hsize}{!}{\includegraphics{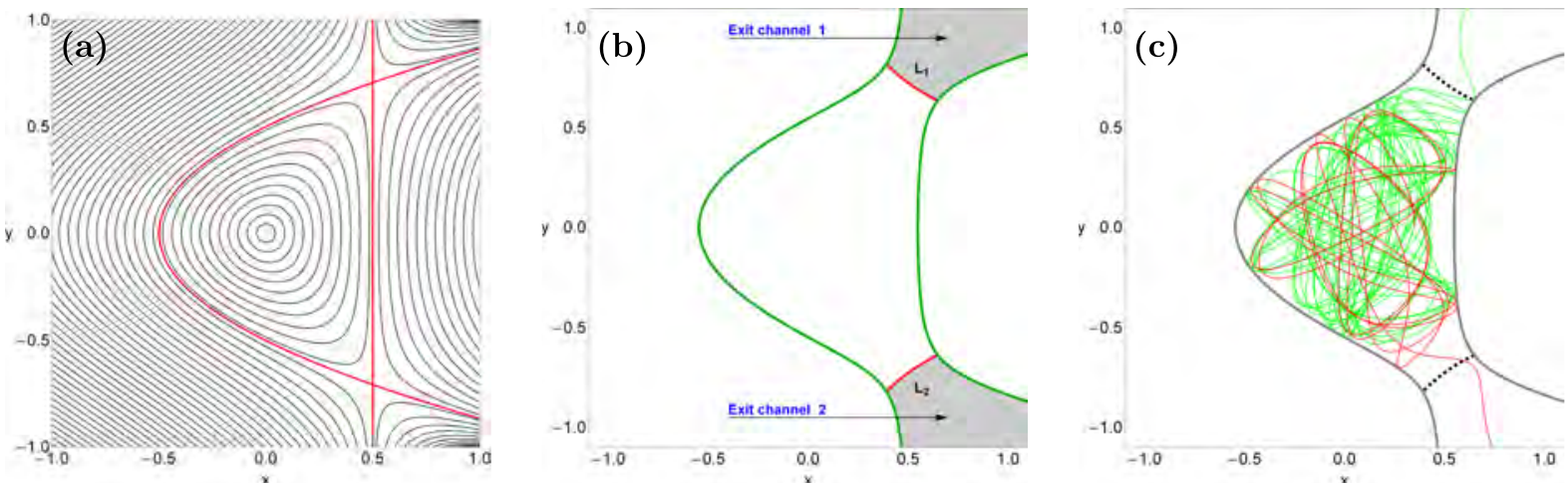}}
\caption{(a): Equipotential curves of the potential (\ref{pot}) for various values of the energy $h$ when $V_1(x,y) = - x y^2$. The equipotential curve corresponding to the energy of escape is shown with red color; (b): The open ZVC at the physical $(x,y)$ plane when $h = 0.15$. $L_1$ and $L_2$ indicate the two unstable Lyapunov orbits plotted in red; (c): Two escaping orbits when $h = 0.15$. The orbit which escapes from channel 1 is potted with green color, while red color is used for the orbits which escapes through channel 2.}
\label{exit2}
\end{figure*}

In this case, the perturbation term is $V_1(x,y) = - x y^2$ and the corresponding Hamiltonian is
\begin{equation}
H_1 = \frac{1}{2}\left(\dot{x}^2 + \dot{y}^2 + x^2 + y^2\right) - x y^2 = h.
\label{ham1}
\end{equation}
This Hamiltonian system has an escape energy which equals to 1/8 and it has been studied extensively in numerous previous papers (e.g., [\citealp{C90}, \citealp{CE04}, \citealp{CHLG12}, \citealp{KSCD99}, \citealp{SCK95}]). This dynamical system has a special symmetry; $H_1$ is symmetric with respect to $y \rightarrow - y$. The equipotential curves of the potential (\ref{pot}) for various values of the energy $h$ are shown in Fig. \ref{exit2}a. The equipotential corresponding to the energy of escape $h_{esc}$ is plotted with red color in the same plot. The open ZVC at the physical $(x,y)$ plane when $h = 0.15 > h_{esc}$ is presented with green color in Fig. \ref{exit2}b and the two channels of escape are shown. In the same plot, we denote the two unstable Lyapunov orbits by $L_1$ and $L_2$ using red color. In Fig. \ref{exit2}c we depict with different colors two orbits, one escaping from channel 1 and the other from channel 2, when $h = 0.15$.

\begin{figure*}[!tH]
\centering
\resizebox{0.90\hsize}{!}{\includegraphics{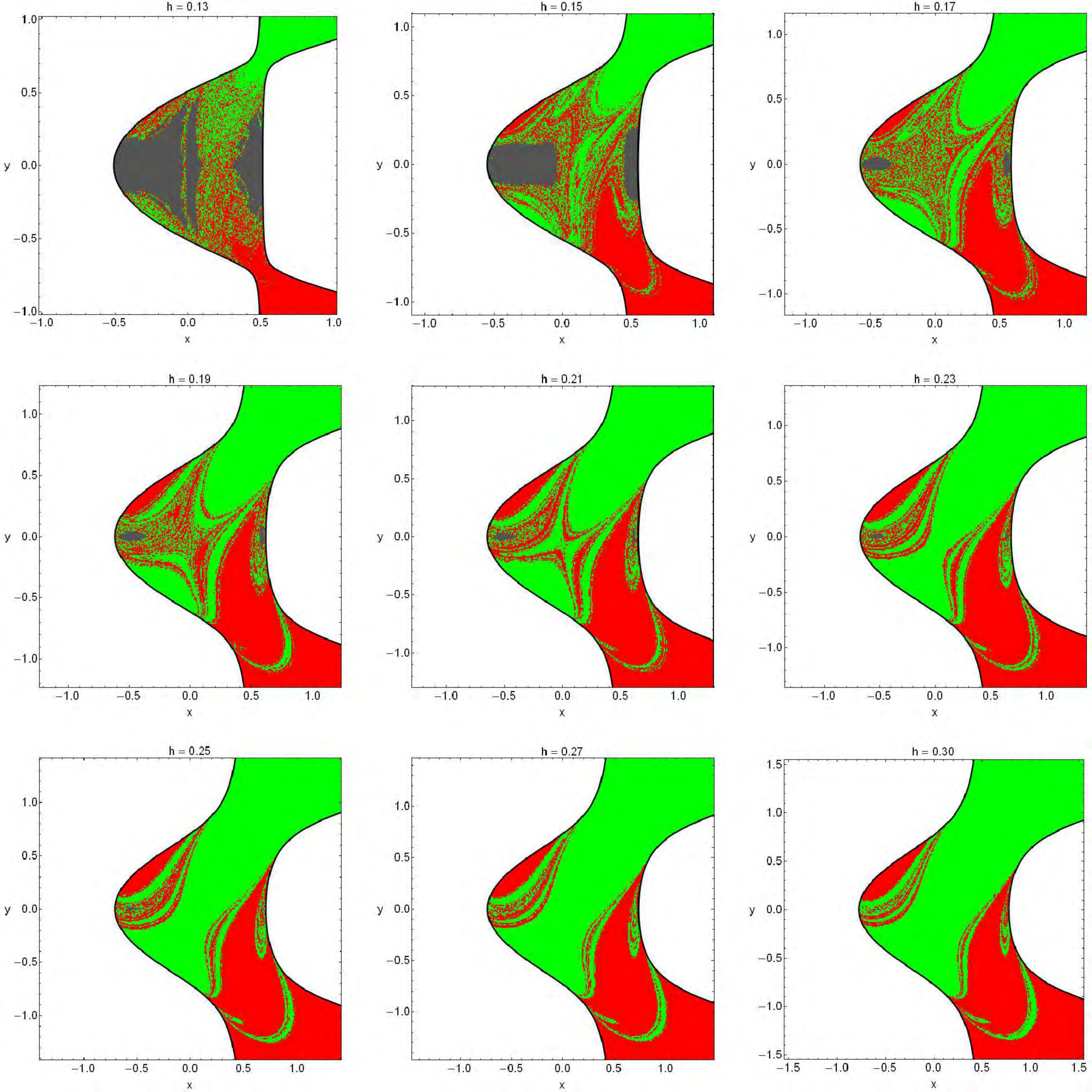}}
\caption{The structure of the physical $(x,y)$ plane for several values of the energy $h$, distinguishing between different escape channels. The color code is as follows: Trapped (gray); escape through channel 1 (green); escape through channel 2 (red).}
\label{cxy2}
\end{figure*}

\begin{figure*}[!tH]
\centering
\resizebox{0.8\hsize}{!}{\includegraphics{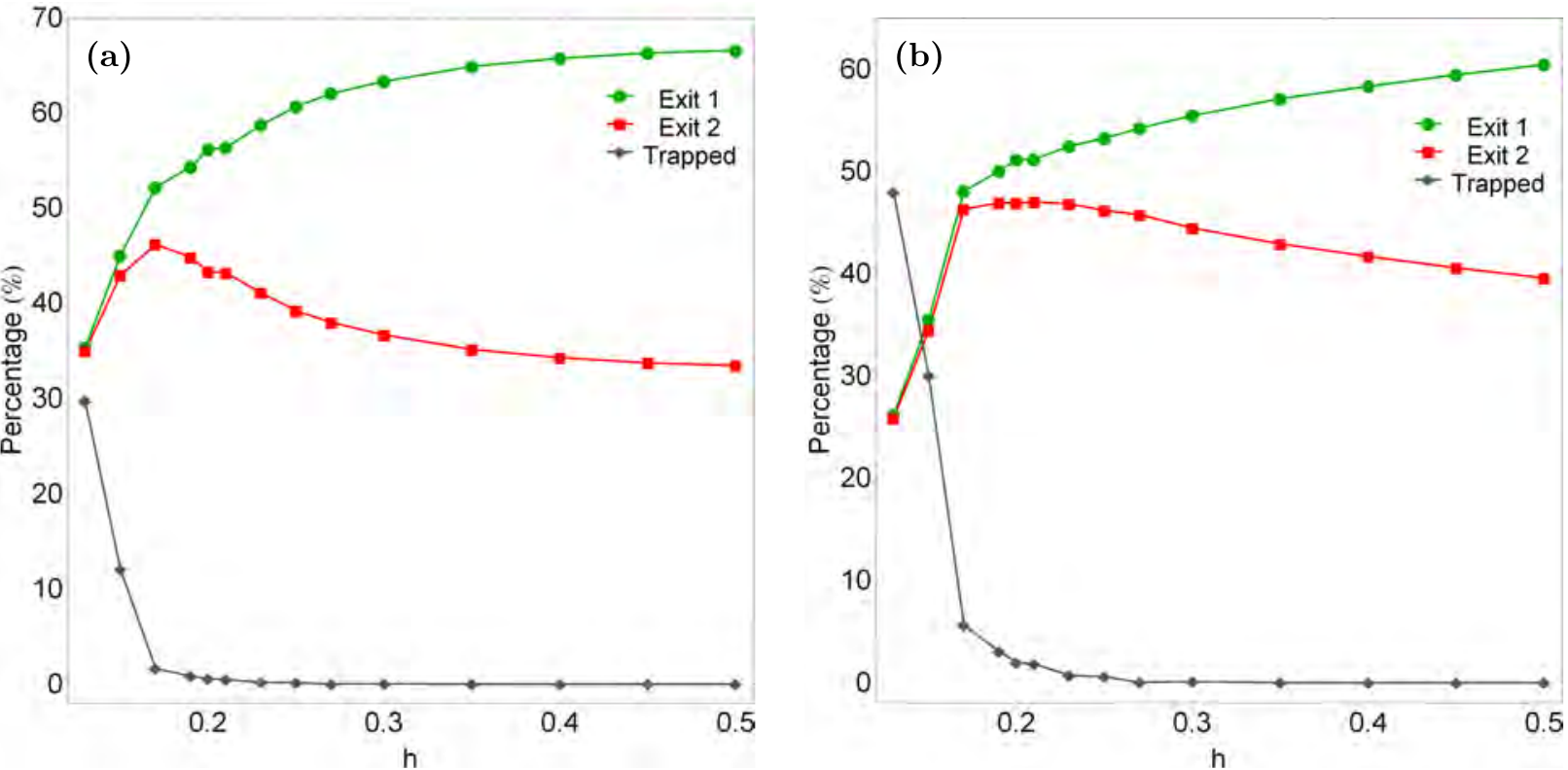}}
\caption{Evolution of the percentages of trapped and escaping orbits when varying the energy $h$ (a-left): on the physical $(x,y)$ plane and (b-right): on the phase $(x,\dot{x})$ plane.}
\label{percs2}
\end{figure*}

\begin{figure*}[!tH]
\centering
\resizebox{0.95\hsize}{!}{\includegraphics{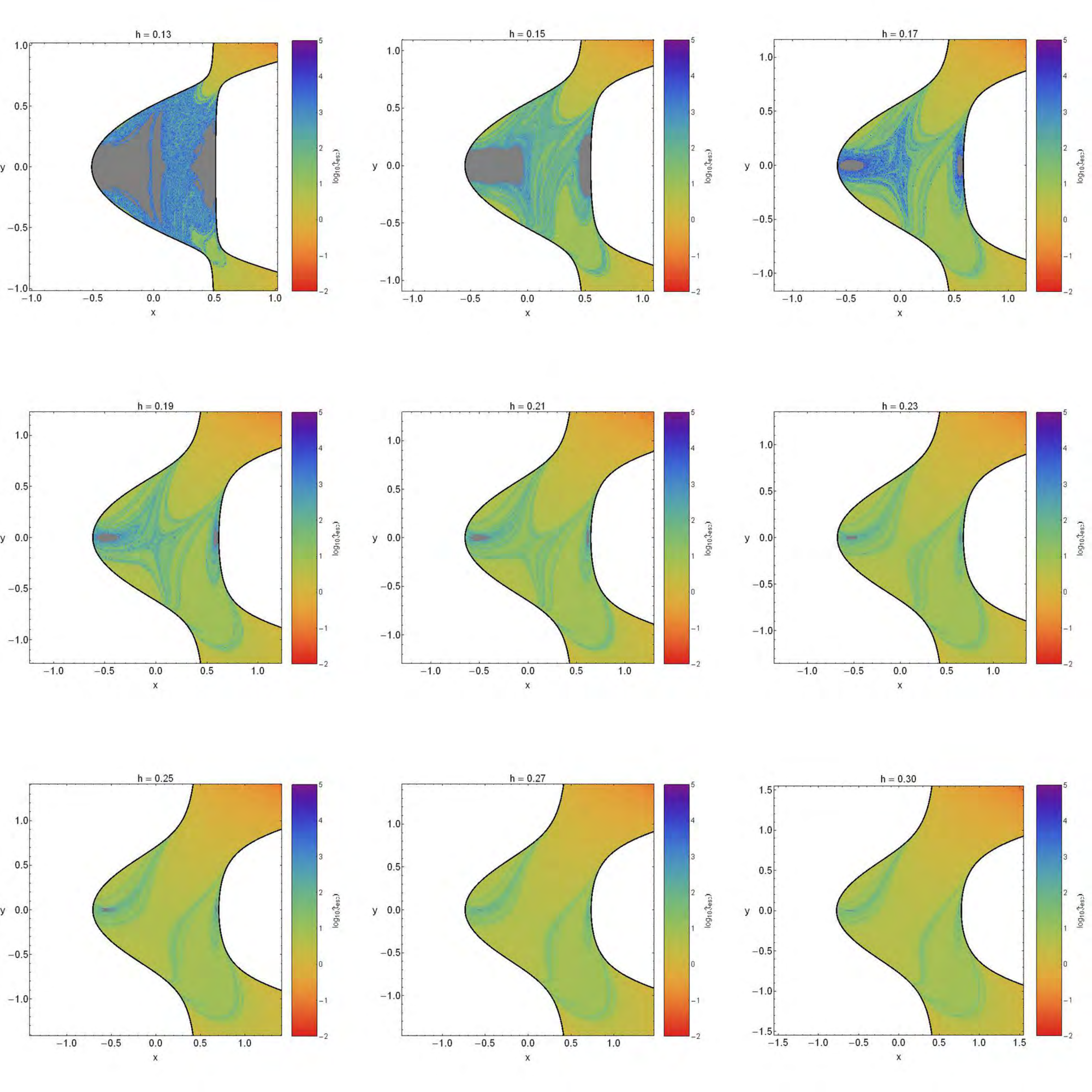}}
\caption{Distribution of the escape times $t_{\rm esc}$ of the orbits on the $(x,y)$ plane. The darker the color, the larger the escape time. Trapped orbits are indicated by gray color.}
\label{txy2}
\end{figure*}

\begin{figure*}[!tH]
\centering
\resizebox{0.90\hsize}{!}{\includegraphics{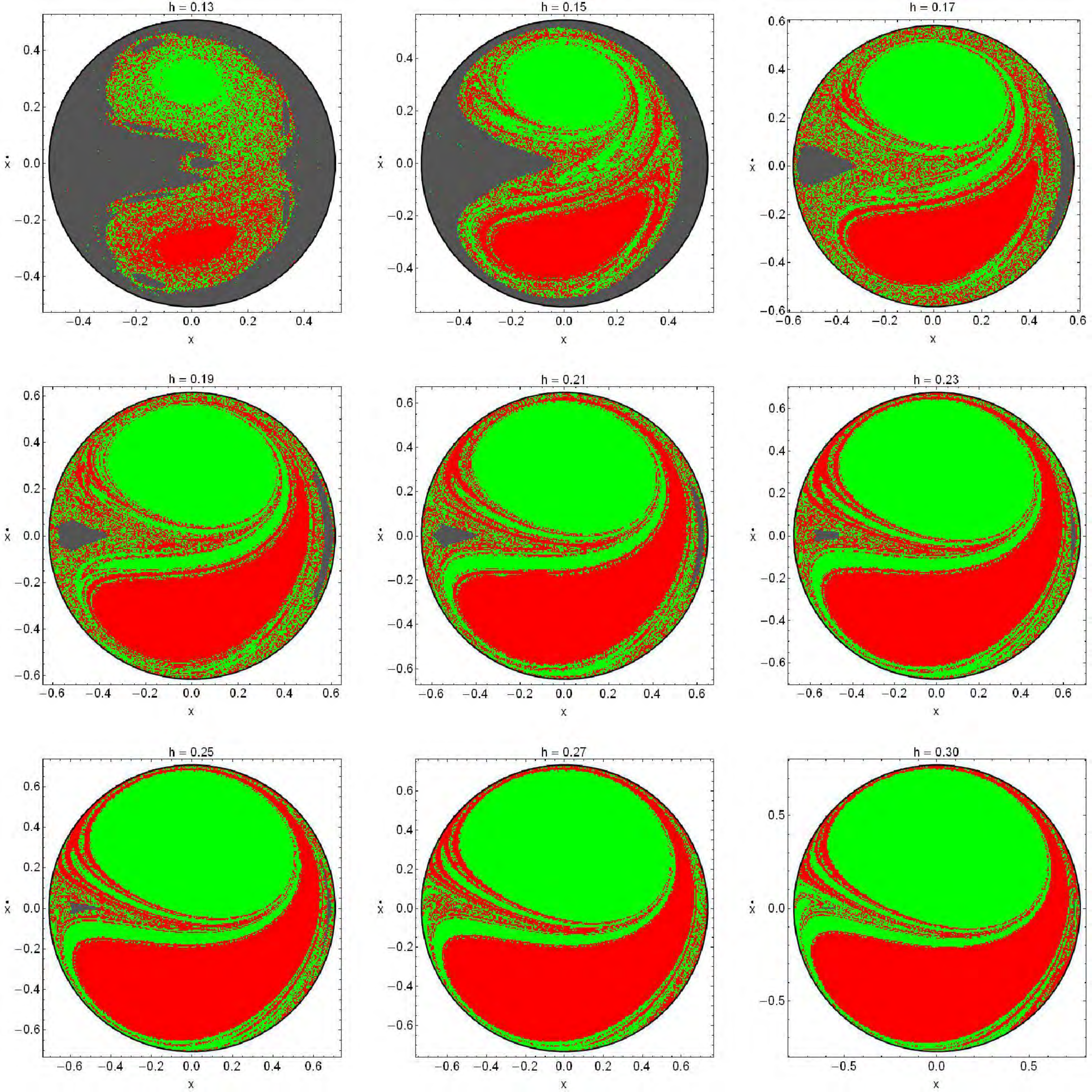}}
\caption{The structure of the phase $(x,\dot{x})$ plane for several values of the energy $h$, distinguishing between different escape channels. The color code is as follows: Trapped (gray); escape through channel 1 (green); escape through channel 2 (red).}
\label{cxpx2}
\end{figure*}

\begin{figure*}[!tH]
\centering
\resizebox{0.95\hsize}{!}{\includegraphics{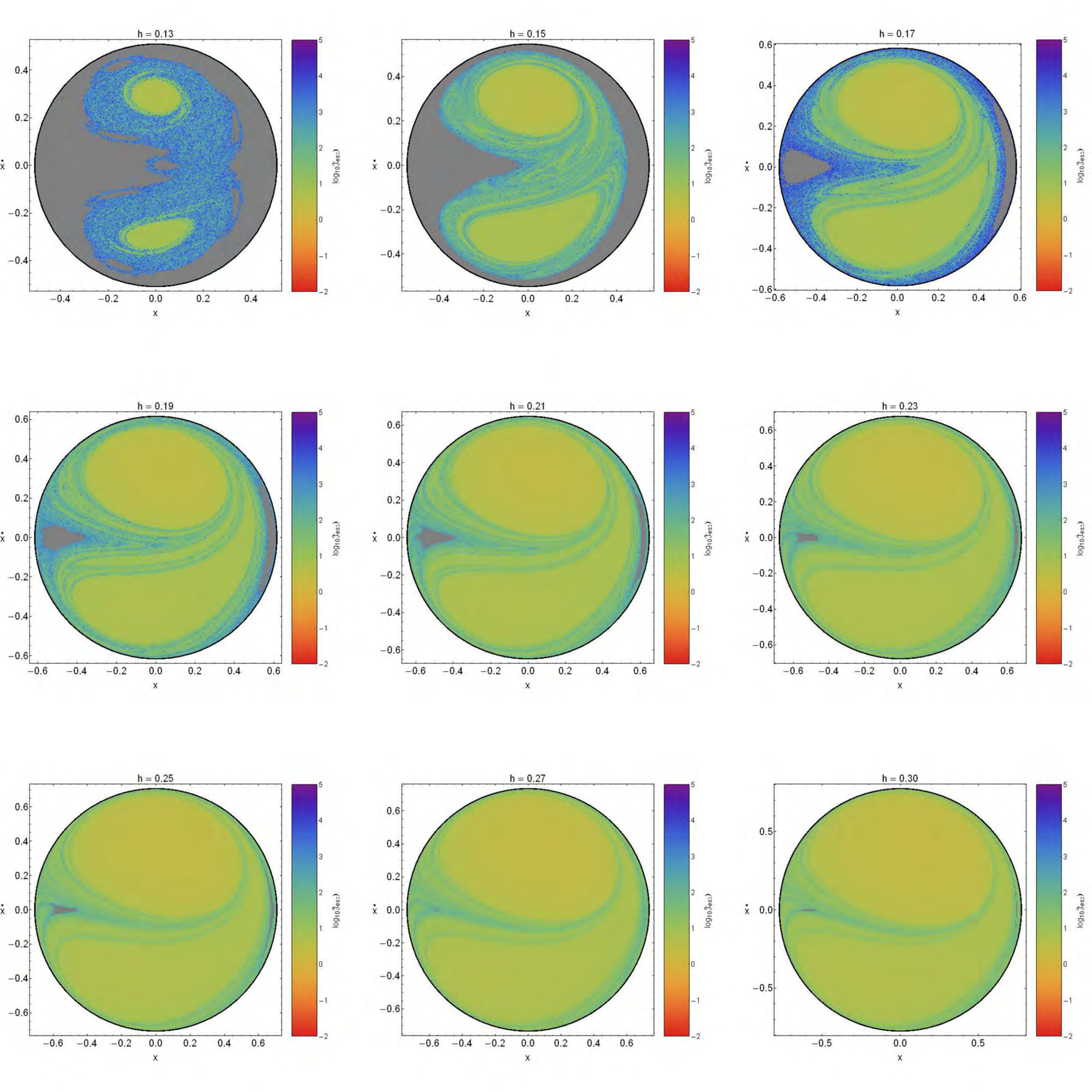}}
\caption{Distribution of the escape times $t_{\rm esc}$ of the orbits on the $(x,\dot{x})$ plane. The darker the color, the larger the escape time. Trapped orbits are indicated by gray color.}
\label{txpx2}
\end{figure*}

\begin{figure*}[!tH]
\centering
\resizebox{0.90\hsize}{!}{\includegraphics{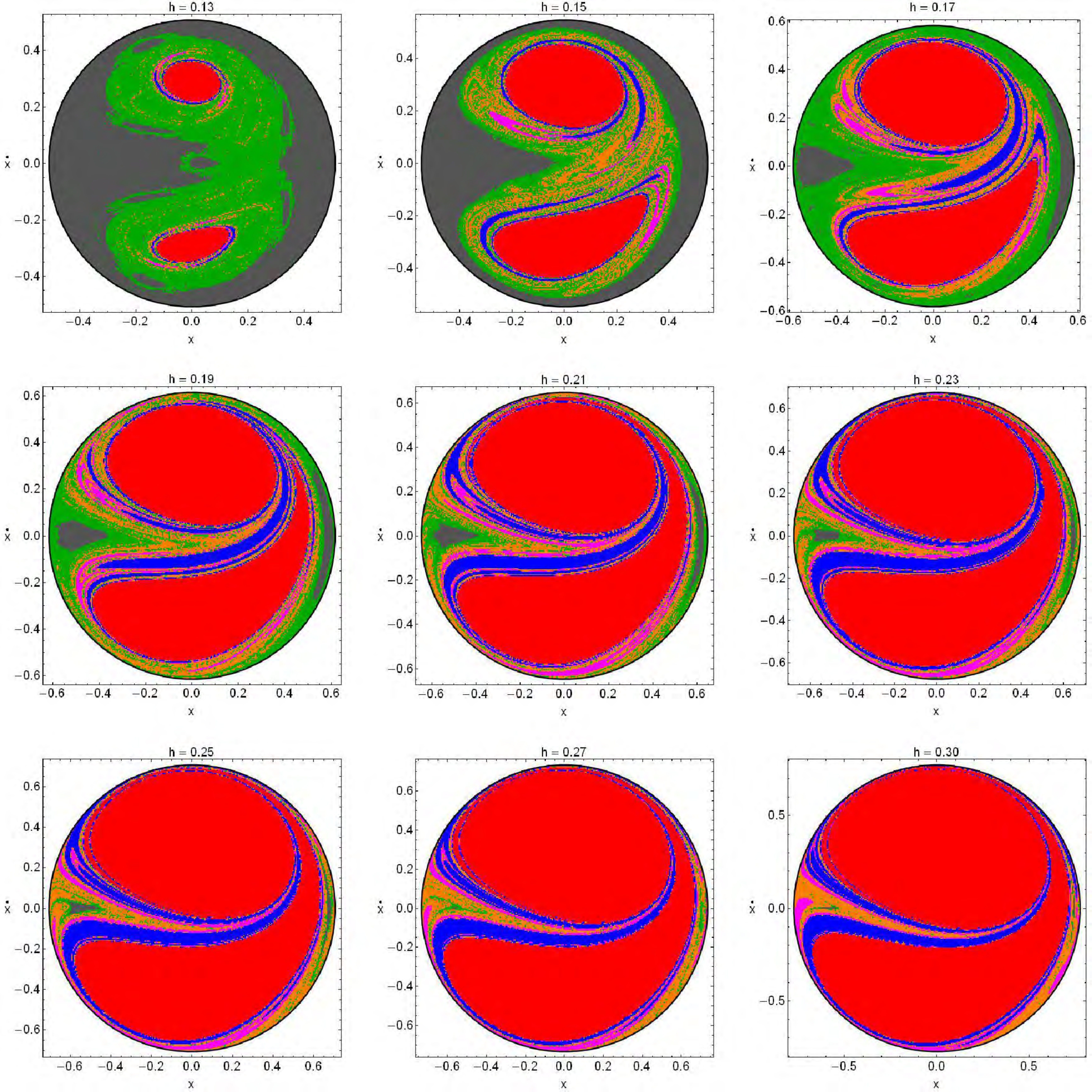}}
\caption{Color scale of the escape regions as a function of the number of intersections with the $y = 0$ axis upwards $(\dot{y} > 0)$. The color code is as follows: 0 intersections (red); 1 intersection (blue); 2 intersections (magenta); 3--10 intersections (orange); $> 10$ intersections (green). The gray regions represent stability islands of trapped orbits.}
\label{inter2}
\end{figure*}

\begin{figure}[!tH]
\includegraphics[width=\hsize]{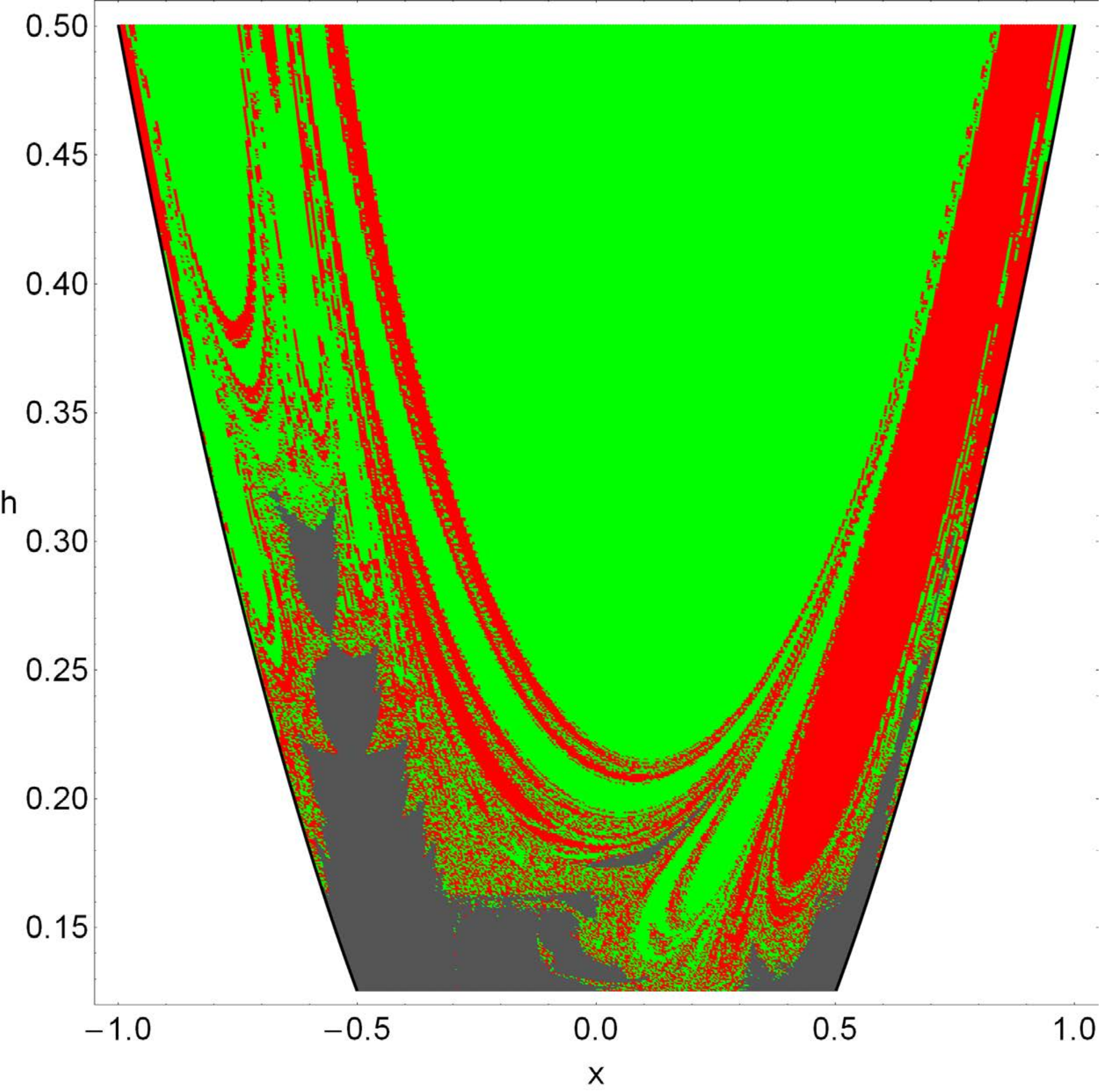}
\caption{Orbital structure of the $(x,h)$-plane when two channels of escape are present. This diagram gives a detailed analysis of the evolution of the trapped and escaping orbits of the dynamical system when the parameter $h$ changes. The color code is as in Fig. \ref{cxy2}.}
\label{xh2}
\end{figure}

In our investigation, we shall deal only with unbounded motion of test particles for values of energy in the set $h = \{0.13, 0.15, 0.17, 0.19, 0.21, 0.23, 0.25, 0.27, 0.30\}$. First of all, we will explore the escape process in the physical $(x,y)$ plane. Fig. \ref{cxy2} shows the structure of the $(x,y)$ plane for different values of the energy. Each initial condition is colored according to the escape channel through which the particular orbit escapes. The gray regions on the other hand, denote initial conditions where the test particles do not escape. The outermost black solid line is the Zero Velocity Curve (limiting curve) which is defined as $V(x,y) = h$. It is seen, that for values of energy larger but yet very close to the escape energy $(h < 0.16)$ a large portion of the $(x,y)$ plane is covered by stability islands which correspond to initial conditions of trapped orbits surrounding by a very rich fractal structure. Looking carefully the grids we also observe that there is a highly sensitive dependence of the escape process on the initial conditions, that is, a slight change in the initial conditions makes the test particle escape through another channel, which is is a classical indication of chaos. As the value of the energy increases the stability islands with trapped regular orbits are reduced and basins of escape emerge. Indeed, when $h = 0.30$ all the computed orbits of the grid escape and there is no indication of bounded motion or whatsoever. By the term basin of escape, we refer to a set of initial conditions that corresponds to a certain escape channel. The escape basins become smoother and more well-defined as the energy increases and the degree of fractility decreases\footnote{The fat-fractal exponent increases, approaching the value 1 which means no fractal geometry, when the energy of the system is high enough (see [\citealp{BBS08}]).}. The fractility is strongly related with the unpredictability in the evolution of a dynamical system. In our case, it can be interpreted that for high enough energy levels, the test particles escape very fast from the scattering region and therefore, the system's predictability increases.

Fig. \ref{percs2}a shows the evolution of the percentages of trapped and escaping orbits on the physical $(x,y)$ plane when the value of the energy $h$ varies. One may observes, that when $h = 0.13$, that is just above the escape energy, trapped, escaping through channel 1 and escaping through channel 2 orbits almost share the entire plane. As the value of the energy increases however, the rate of trapped orbits drops rapidly and when $h > 0.28$ it vanishes. At the same time, the percentage of orbits escaping through channel 1 increases steadily and for $h > 0.35$, it seems to saturate around 65\%, thus occupying around two thirds of the $(x,y)$ plane. On the other hand, the rate of orbits escaping through channel 2 increases for $h < 0.17$ but then is exhibits a slow reduction and for $h > 0.37$ it saturates around 35\%. Therefore, one may concludes that for high energy levels $(h > 0.35)$, all orbits in the $(x,y)$ plane escape and about two thirds of them choose channel 1.

The following Fig. \ref{txy2} shows how the escape times $t_{\rm esc}$ of orbits are distributed on the $(x,y)$ plane. Light reddish colors correspond to fast escaping orbits, dark blue/purpe colors indicate large escape periods, while gray color denote trapped orbits. We observe, that when $h = 0.13$, that is a value of energy very close to the escape energy, the escape periods of the majority of orbits are huge corresponding to tens of thousands of time units. This however, is anticipated because in this case the width of the escape channels is very small and therefore, the orbits should spend much time inside the equipotential curve until they find one of the openings and eventually escape to infinity. As the value of the energy increases however, the escape channels become more and more wide leading to faster escaping orbits, which means that the escape period decreases rapidly. We found, that the longest escape rates correspond to initial conditions near the boundaries between the escape basins and near the vicinity of stability islands. On the other hand, the shortest escape periods have been measured for the regions without sensitive dependence on the initial conditions (basins of escape), that is, those far away from the fractal basin boundaries.

We continue our exploration of the escape process in the phase $(x,\dot{x})$ plane. The structure of the $(x,\dot{x})$ phase plane for several values of the energy is shown in Fig. \ref{cxpx2}. We observe a similar behavior to that discussed for the physical $(x,y)$ plane in Fig. \ref{cxy2}. The outermost black solid line is the Zero Velocity Curve (limiting curve) which is defined as
\begin{equation}
f(x,\dot{x}) = \frac{1}{2}\dot{x}^2 + V(x, y = 0) = h.
\label{zvc}
\end{equation}
It is worth noticing, that in the phase plane the limiting curve is closed but this does not mean that there is no escape. Remember, that we decided to choose such perturbation terms that produce the escape channels on the physical $(x,y)$ plane which is a subspace of the entire four-dimensional $(x,y,\dot{x},\dot{y})$ space of the system. Here we must point out, that this $(x,\dot{x})$ phase plane is not a Poincar\'{e} Surface of Section (PSS), simply because escaping orbits in general, do not intersect the $y = 0$ axis after a certain time, thus preventing us from defying a recurrent time. A classical Poincar\'{e} surface of section exists only if orbits intersect an axis like $y = 0$ at least once within a certain time interval. Nevertheless, in the case of escaping orbits we can still define local surfaces of section which help us to understand the orbital behavior of the dynamical system.

Again, we can distinguish in the phase plane fractal regions where we cannot predict the particular escape channel and regions occupied by escape basins. These basins are either broad well-defined regions, or elongated bands of complicated structure spiralling around the center. We see that again for values of energy close to the escape energy there is a considerable amount of trapped orbits and the degree of fractalization of the phase plane is high. As we proceed to higher energy levels however, the rate of trapped orbits reduces, the phase plane becomes less and less fractal and is occupied by well-defined basins of escape. In Fig. \ref{percs2}b we present the evolution of the percentages of trapped and escaping orbits on the phase plane when the value of the energy $h$ varies. It is observed, that the pattern and the evolution of the percentages is almost identical to that discussed in Fig. \ref{percs2}a regarding the physical plane. In particular, for $h = 0.13$ about half of the phase plane is covered by initial conditions corresponding to trapped orbits, while the escaping orbits share the rest half of the $(x,\dot{x})$ plane. At the highest energy level studied $(h = 0.5)$, about 60\% of the total orbits escape through channel 1 and 40\% through channel 2; the percentage of trapped orbits has already reached the zero value from $h > 0.3$.

The distribution of the escape times $t_{\rm esc}$ of orbits on the $(x,\dot{x})$ plane is shown in Fig. \ref{txpx2}. It is evident, that orbits with initial conditions inside the exit basins escape from the system very quickly, or in other words, they possess extremely small escape periods. On the contrary, orbits with initial conditions located in the fractal parts of the phase plane need considerable amount of time in order to escape. Another interesting way of measuring the escape rate of an orbit is by counting how many intersection the orbit has with the axis $y = 0$ before it escapes. The regions in Fig. \ref{inter2} are colored according to the number of intersections with the axis $y = 0$ upwards $(\dot{y} > 0)$. We observe, that orbits with initial conditions inside the two red basins escape directly without any intersection with the $y = 0$ axis. Furthermore, as the value of the energy increases, these red regions grow in relative size (proportion of the total area on the phase plane) and for high enough energy levels they occupy around 90\% of the grid. We should also note, that orbits with initial conditions located at the vicinity of the stability islands perform numerous intersections with the $y = 0$ axis before they eventually escape to infinity. On the other hand, orbits with initial conditions in the elongated spiral bands need only a couple of intersection until they escape.

The grids in physical $(x,y)$ as well as the phase $(x,\dot{x})$ plane provide information on the phase space mixing for only a fixed value of energy. H\'{e}non however, back in the 60s, considered a plane which provides information about regions of stability and regions of escaping orbits using the section $y = \dot{x} = 0$, $\dot{y} > 0$, i.e., the test particle starts on the $x$-axis, parallel to the $y$-axis and in the positive $y$-direction. Thus, in contrast to the previously discussed grids, only orbits with pericenters on the $x$-axis are included and therefore, the value of the energy $h$ is used as an ordinate. Fig. \ref{xh2} shows the structure of the $(x,h)$-plane when $h \in (0.125, 0.5]$. The boundaries between bounded and unbounded motion are now seen to be more jagged than shown in the previous grids. In addition, we found in the blow-ups of the diagram many tiny islands of stability\footnote{From chaos theory we expect an infinite number of islands of (stable) quasi-periodic (or small scale chaotic) motion.}. We see, that for low values of the energy close to the escape energy, there is a considerable amount of trapped orbits inside stability regions surrounding by a highly fractal structure. This pattern however changes for larger energy levels, where there are no trapped orbits and the vast majority of the grid is covered by well-formed basins of escape, while fractal structure is confined only near the boundaries of the escape basins.

\begin{figure*}[!tH]
\centering
\resizebox{0.80\hsize}{!}{\includegraphics{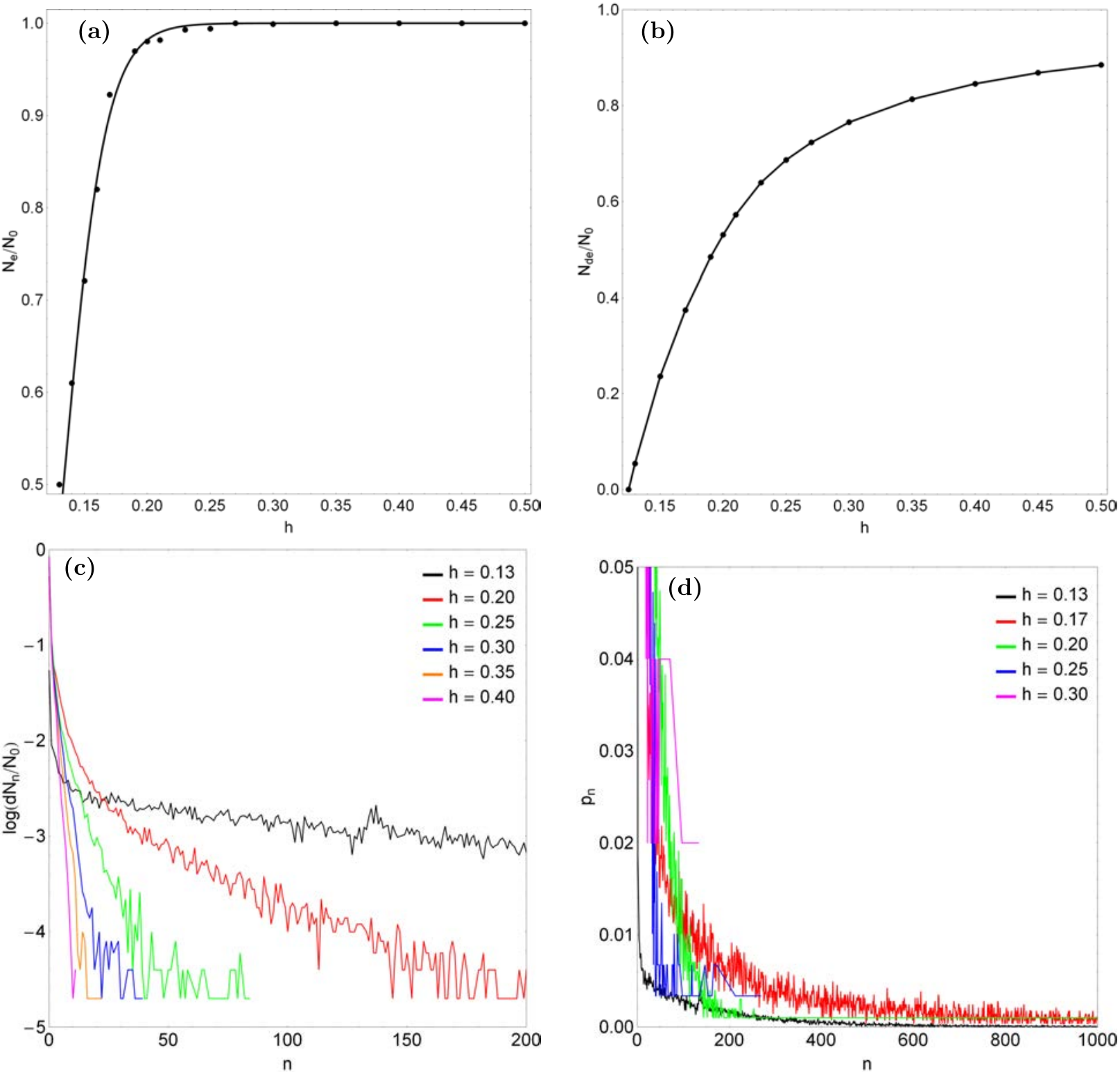}}
\caption{(a-upper left): Evolution of the proportion of escaping orbits $N_e/N_0$ as a function of the energy $h$, (b-upper right): Evolution of the proportion of directly escaping orbits $N_{de}/N_0$ as a function of the energy $h$, (c-lower left): Evolution of the logarithmic proportion $dN_n/N_0$ as a function of the number of the intersections $n$, for various values of the energy and (d-lower right): Evolution of the probability $p_n$ of escapes as a function of $n$ for several energy levels.}
\label{gm2}
\end{figure*}

It is of particular interest to conduct a statistical analysis of the escape process in the case of the $(x,\dot{x})$ phase plane. For this purpose, we shall follow the numerical approach used recently in [\citealp{CHLG12}]. Our results are shown in Fig. \ref{gm2}(a-d) where curve fit approximation versus results from numerical integration is presented in (a-b) panels. To begin with, Fig. \ref{gm2}a shows the proportion of escaping orbits $N_e/N_0$ as a function of the energy $h$. For values of energy beyond the escape energy, the majority of orbits escape from the system. Our numerical calculation verify that the evolution of the proportion of escaping orbits can be approximated by the formula
\begin{equation}
N_e/N_0(h) = 0.5\left[1 + \tanh\left(30h - 4\right)\right],
\label{tr21}
\end{equation}
proposed in [\citealp{CHLG12}]. In Fig. \ref{gm2}b we present the evolution of the direct escaping orbits $N_{de}/N_0$ (by the term direct escaping orbits we refer to orbits that escape to infinity immediately without any intersection with the $y = 0$ axis) as a function of the energy $h$. We see, that the amount of direct escaping orbits grows rapidly with increasing $h$ and for high energy levels $(h > 0.5)$ they take over almost all the phase plane (more than 90\%). The proportion of direct escapes can be given by the approximate formula
\begin{equation}
N_{de}/N_0(h) = -1.7 + 19.24 h - 49.15 h^2 + 42.16 h^3.
\label{tr22}
\end{equation}
Moreover, Fig. \ref{gm2}c depicts the logarithm of the proportion of escaping orbits $dN_n/N_0$, where $dN_n$ corresponds to the number of escaping orbits after the $n$th intersection with the $y = 0$ axis upwards $(\dot{y} > 0)$. It is seen, that the escape time of orbits decreases with increasing $n$. In particular, the escape rates are high for relatively small $n$, while they drop rapidly for larger $n$. Last but not least, we computed the probability of escape as a function of the number of intersections for various values of the energy. Specifically, the probability is defined as
\begin{equation}
p_n = \frac{d N_n}{N_n},
\label{tr3}
\end{equation}
where $N_n$ is the number of orbits that have not yet escaped before the $n$th intersection. The evolution of $p_n$ as a function of $n$ for various energy levels is given in Fig. \ref{gm2}d. Here we have to stress out, that the properties of the probability of escape in this system and in other similar systems have been studied in detail in previous papers (e.g., [\citealp{CKK93}, \citealp{SCK95} -- \citealp{SKCD96}]. Furthermore, our numerical calculations regarding the statistical analysis of the dynamical system in the case where two escape channels are present, have found to coincide with the corresponding results given in [\citealp{CHLG12}].

\subsection{Case II: Three channels of escape}
\label{case2}

\begin{figure*}[!tH]
\centering
\resizebox{\hsize}{!}{\includegraphics{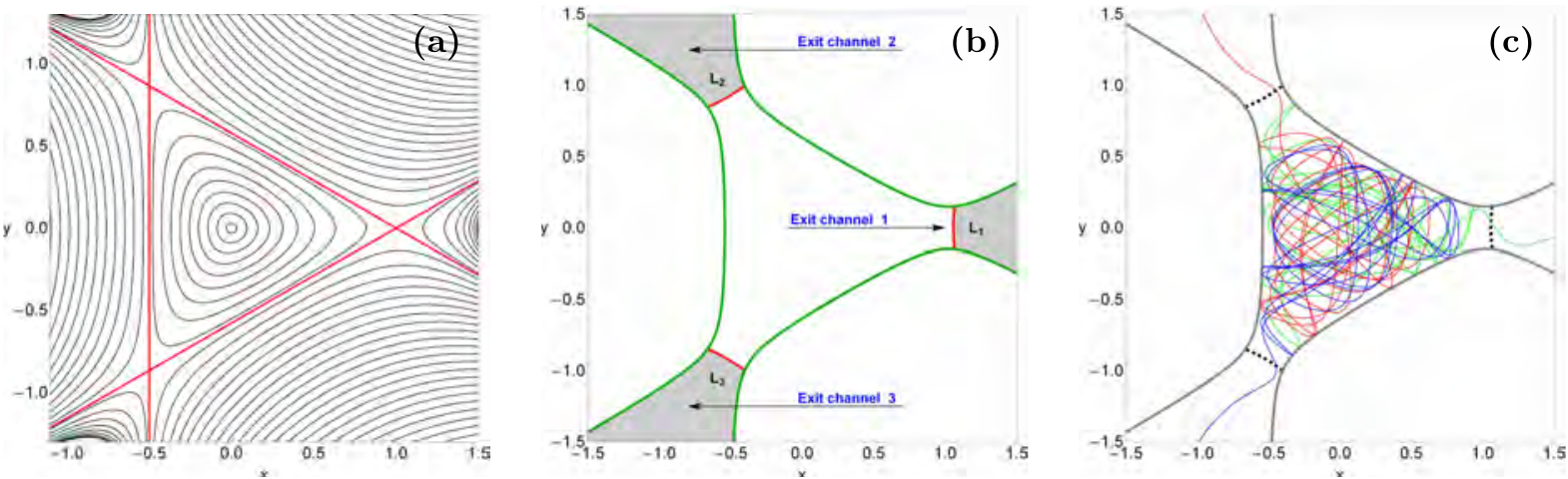}}
\caption{(a): Equipotential curves of the potential (\ref{pot}) for various values of the energy $h$ when $V_1(x,y) = - x \left(x^2/3 - y^2\right)$. The equipotential curve corresponding to the energy of escape is shown with red color; (b): The open ZVC at the physical $(x,y)$ plane when $h = 0.2$. $L_1$, $L_2$ and $L_3$ indicate the three unstable Lyapunov orbits plotted in red; (c): Three escaping orbits when $h = 0.2$. The orbit which escapes from channel 1 is potted with green color, the orbit escaping from channel 2 with red color, while blue color is used for the orbits which escapes through channel 3.}
\label{exit3}
\end{figure*}

We continue our exploration of escapes in a Hamiltonian system with three exit channels and escape energy equal to 1/6. In order to obtain this number of exits in the limiting curve in the $(x,y)$ plane, the perturbation term should be $V_1(x,y) = - x \left(x^2/3 - y^2\right)$ and the corresponding Hamiltonian reads
\begin{equation}
H_2 = \frac{1}{2}\left(\dot{x}^2 + \dot{y}^2 + x^2 + y^2\right) - x \left(x^2/3 - y^2\right) = h.
\label{ham2}
\end{equation}
$H_2$ manifests a $2\pi/3$ rotation symmetry, but for $\varepsilon$ this discrete symmetry is broken. Here we should like to note, that the particular type of the perturbation is very similar to that of the classical H\'{e}non-Heiles Hamiltonian system [\citealp{HH64}] (in fact we changed the position of the $x$ and $y$ variables). We made this choice mainly for two reasons: (i) the standard H\'{e}non-Heiles dynamical system has been studied extensively and thoroughly in numerous papers over the last years (e.g., [\citealp{AVS01} -- \citealp{AVS03}, \citealp{BBS08}, \citealp{BSBS12}, \citealp{dML99}, \citealp{S07}]) so, we preferred to work on something rather different and (ii) in all cases we wanted the $(x,\dot{x})$ phase plane\footnote{The $(x,\dot{x})$ phase plane is constructible only if the potential has terms with even powers regarding the $y$ variable.}. It should be pointed out however, that this change in the variables affects only the symmetry, while all the measured quantities remain the same as in the classical H\'{e}non-Heiles system. In Fig. \ref{exit3}a we see the equipotential curves of the potential (\ref{pot}) for various values of the energy $h$, while the equipotential corresponding to the energy of escape $h_{esc}$ is plotted with red color in the same plot. Furthermore, the open ZVC at the physical $(x,y)$ plane when $h = 0.2 > h_{esc}$ is presented with green color in Fig. \ref{exit3}b and the three channels of escape are shown. In the same figure, the three unstable Lyapunov orbits $L_1$, $L_2$ and $L_3$ are denoted using red color. Fig. \ref{exit3}c depicts with different colors three orbits, one escaping from channel 1, one from channel 2 and the other from channel 3, when $h = 0.2$.

\begin{figure*}[!tH]
\centering
\resizebox{0.90\hsize}{!}{\includegraphics{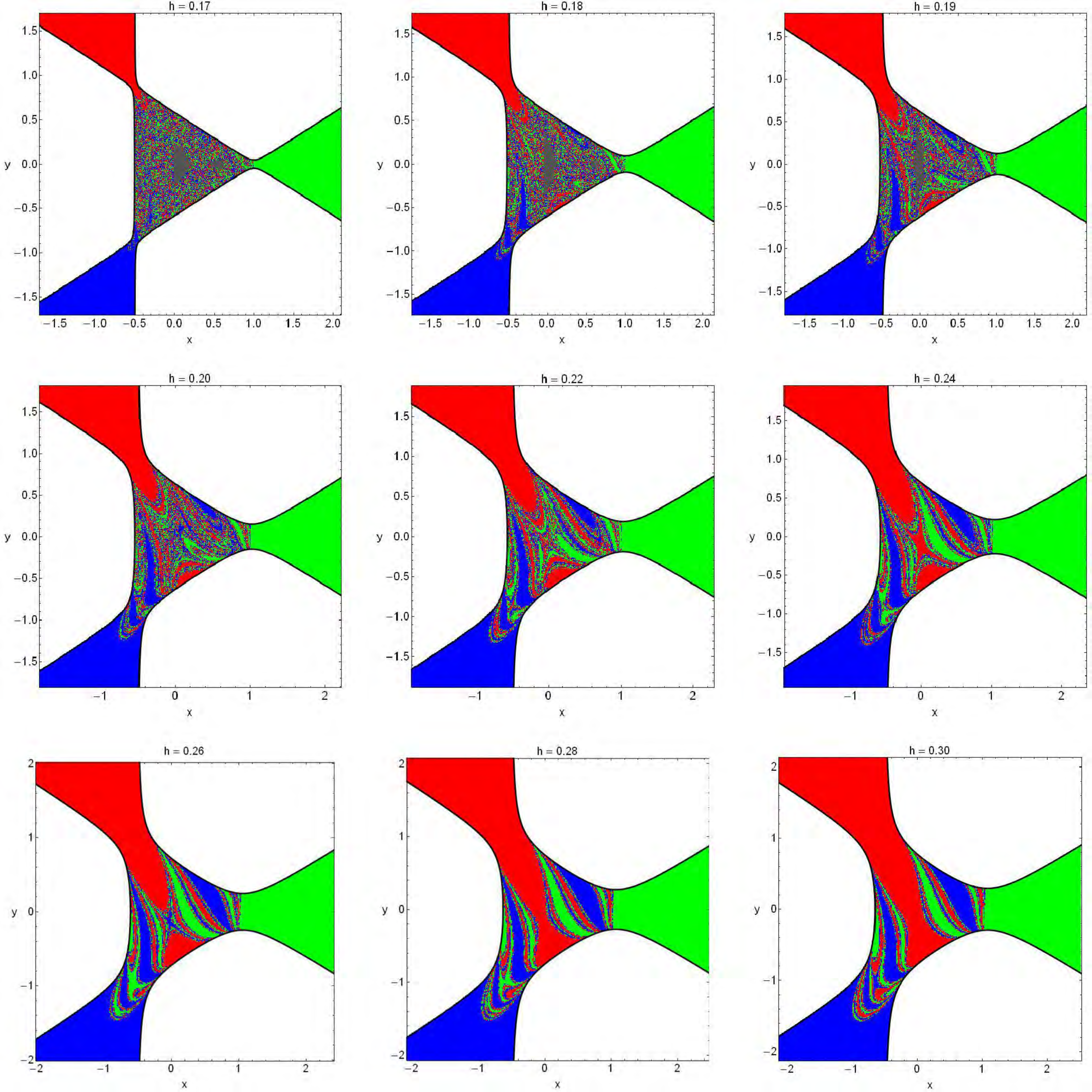}}
\caption{The structure of the physical $(x,y)$ plane for several values of the energy $h$, distinguishing between different escape channels. The color code is as follows: Trapped (gray); escape through channel 1 (green); escape through channel 2 (red); escape through channel 3 (blue).}
\label{cxy3}
\end{figure*}

\begin{figure*}[!tH]
\centering
\resizebox{0.8\hsize}{!}{\includegraphics{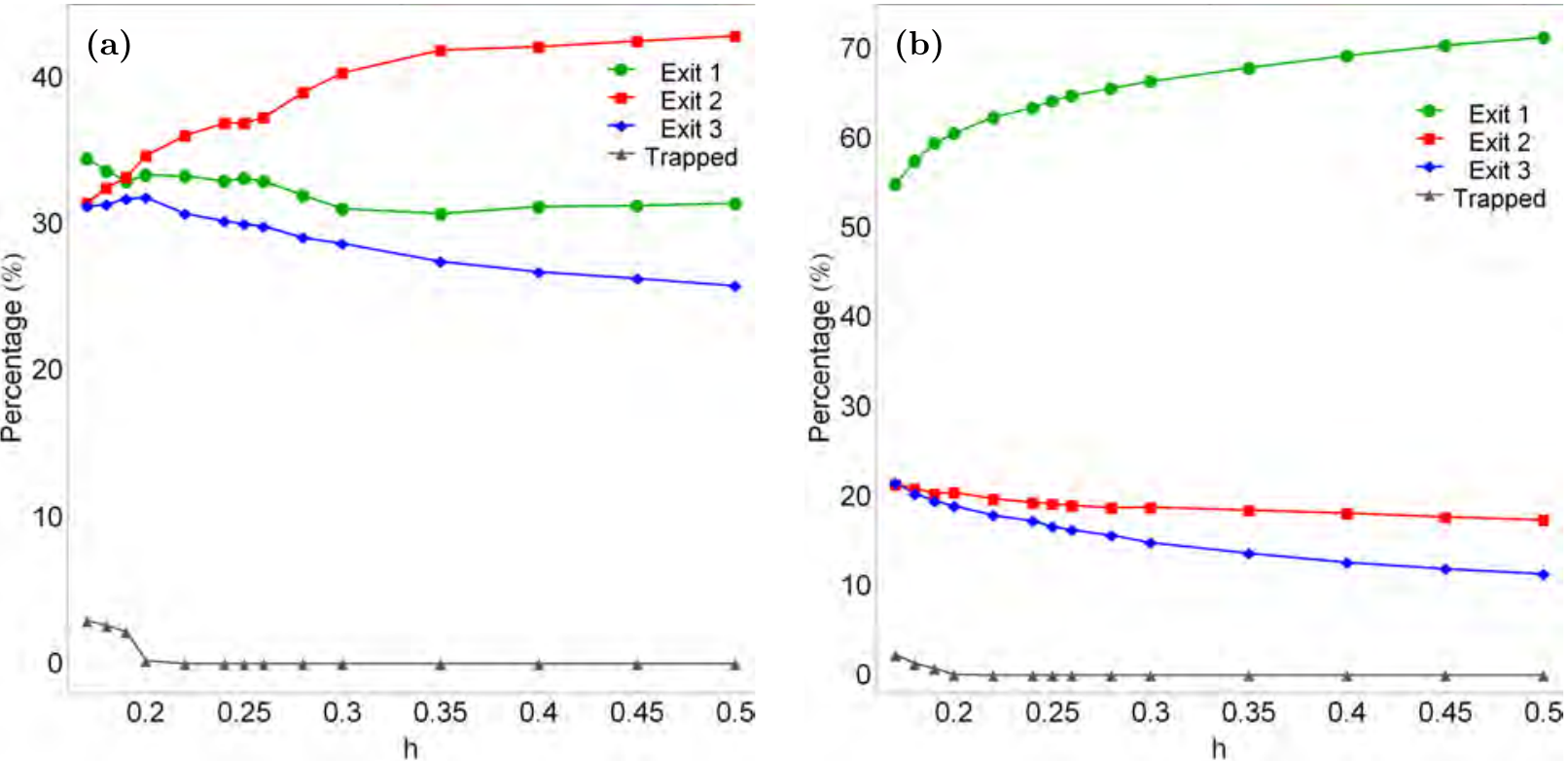}}
\caption{Evolution of the percentages of trapped and escaping orbits when varying the energy $h$ (a-left): on the physical $(x,y)$ plane and (b-right): on the phase $(x,\dot{x})$ plane.}
\label{percs3}
\end{figure*}

\begin{figure*}[!tH]
\centering
\resizebox{0.95\hsize}{!}{\includegraphics{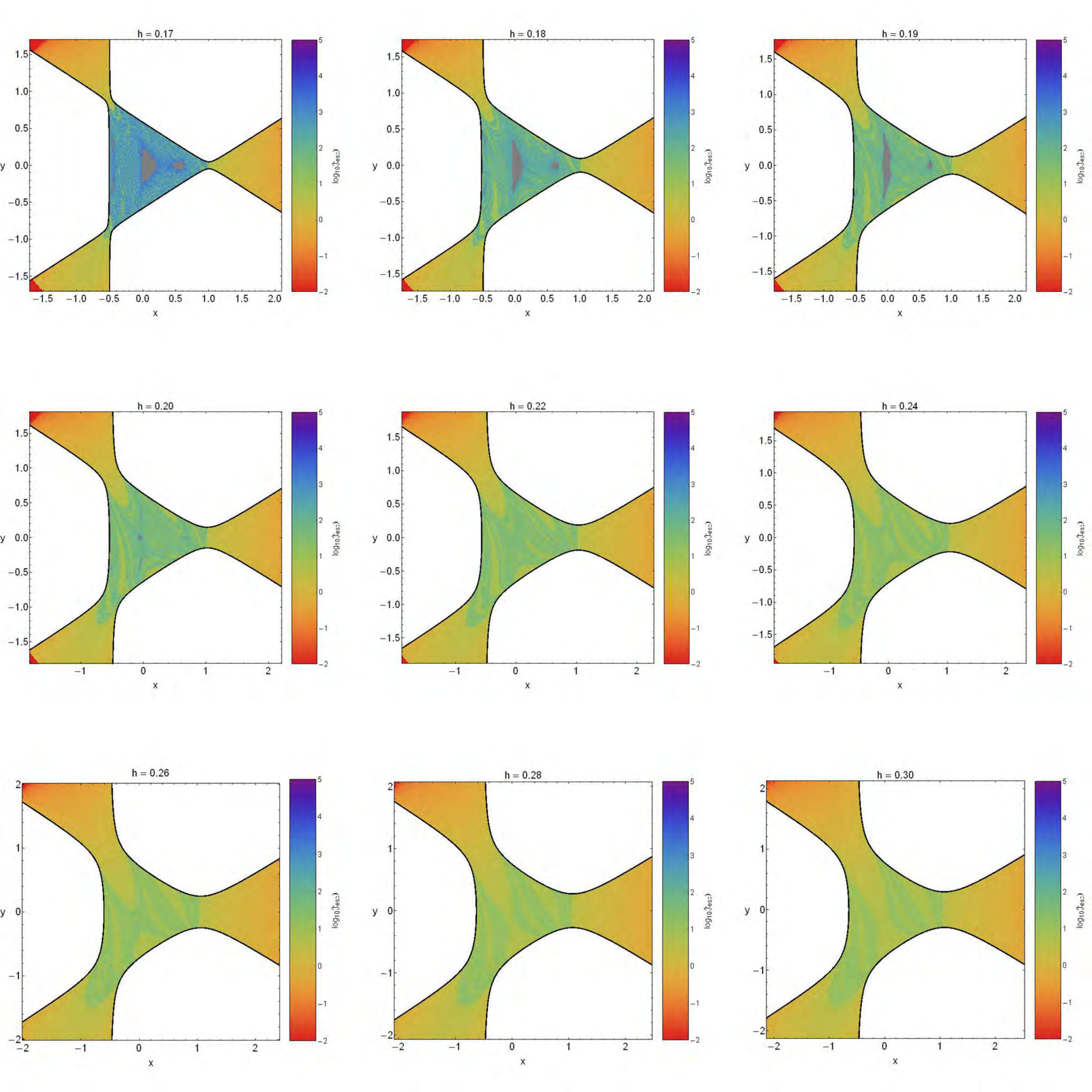}}
\caption{Distribution of the escape times $t_{\rm esc}$ of the orbits on the $(x,y)$ plane. The darker the color, the larger the escape time. Trapped orbits are indicated by gray color.}
\label{txy3}
\end{figure*}

\begin{figure*}[!tH]
\centering
\resizebox{0.90\hsize}{!}{\includegraphics{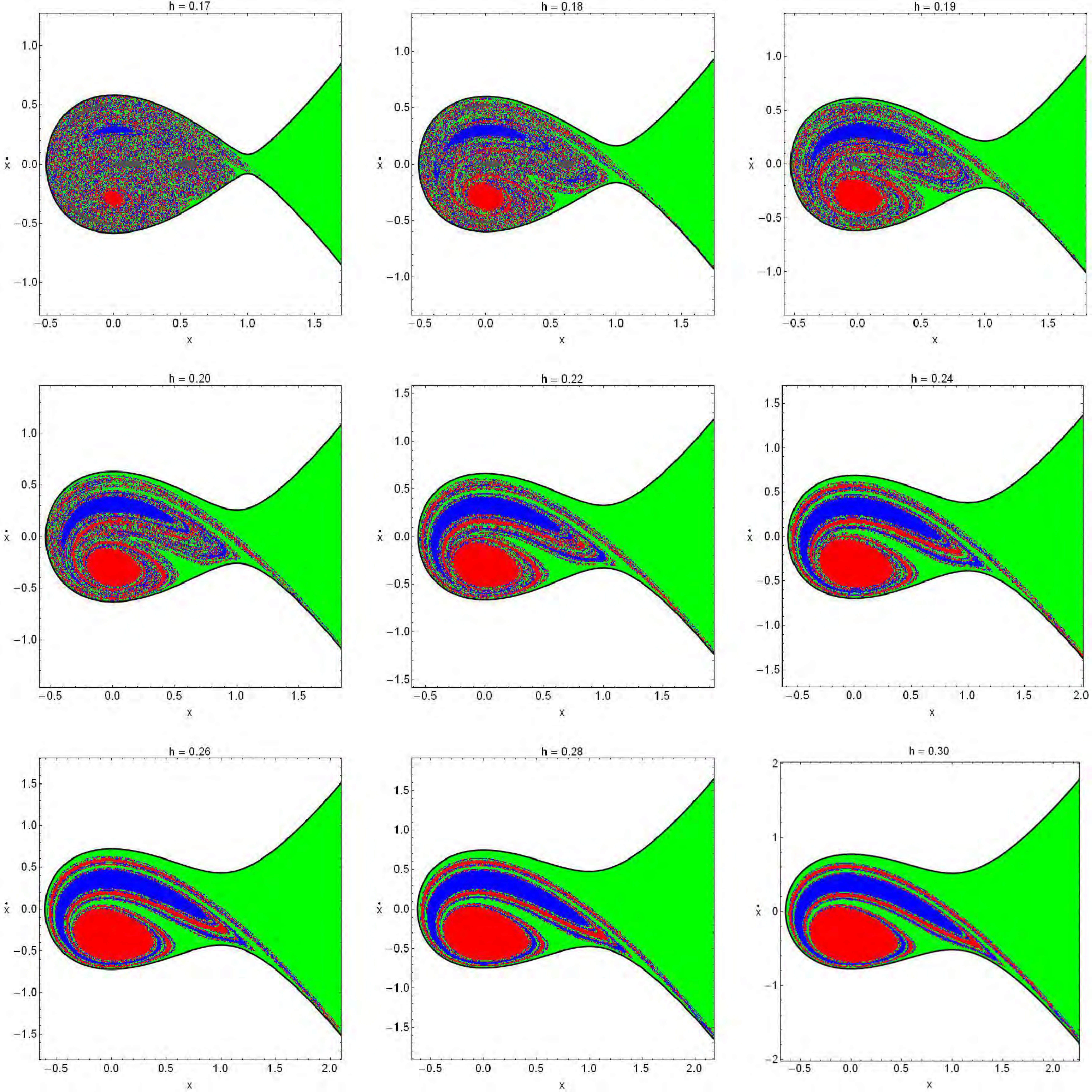}}
\caption{The structure of the phase $(x,\dot{x})$ plane for several values of the energy $h$, distinguishing between different escape channels. The color code is as follows: Trapped (gray); escape through channel 1 (green); escape through channel 2 (red); escape through channel 3 (blue).}
\label{cxpx3}
\end{figure*}

\begin{figure*}[!tH]
\centering
\resizebox{0.95\hsize}{!}{\includegraphics{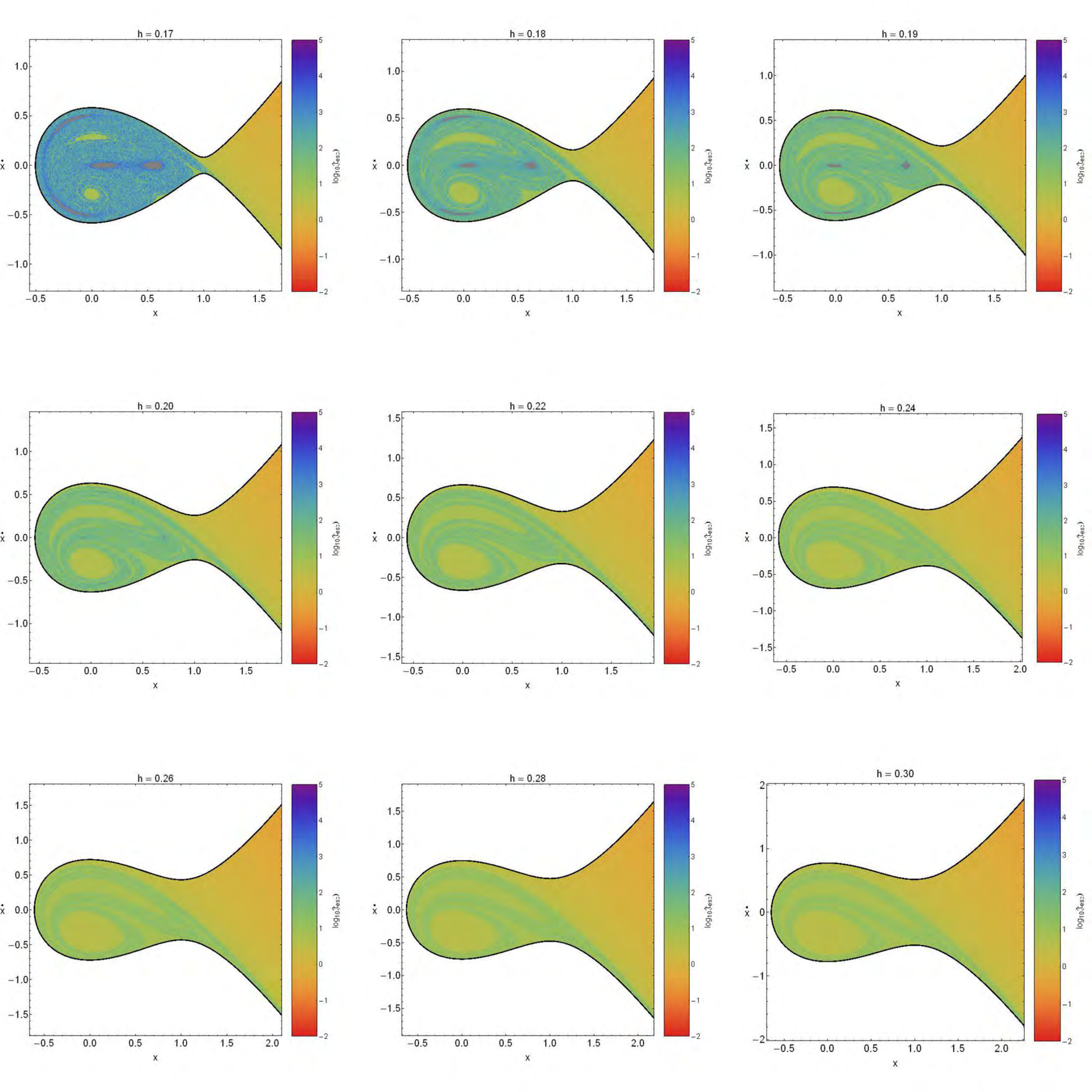}}
\caption{Distribution of the escape times $t_{\rm esc}$ of the orbits on the $(x,\dot{x})$ plane. The darker the color, the larger the escape time. Trapped orbits are indicated by gray color.}
\label{txpx3}
\end{figure*}

\begin{figure*}[!tH]
\centering
\resizebox{0.90\hsize}{!}{\includegraphics{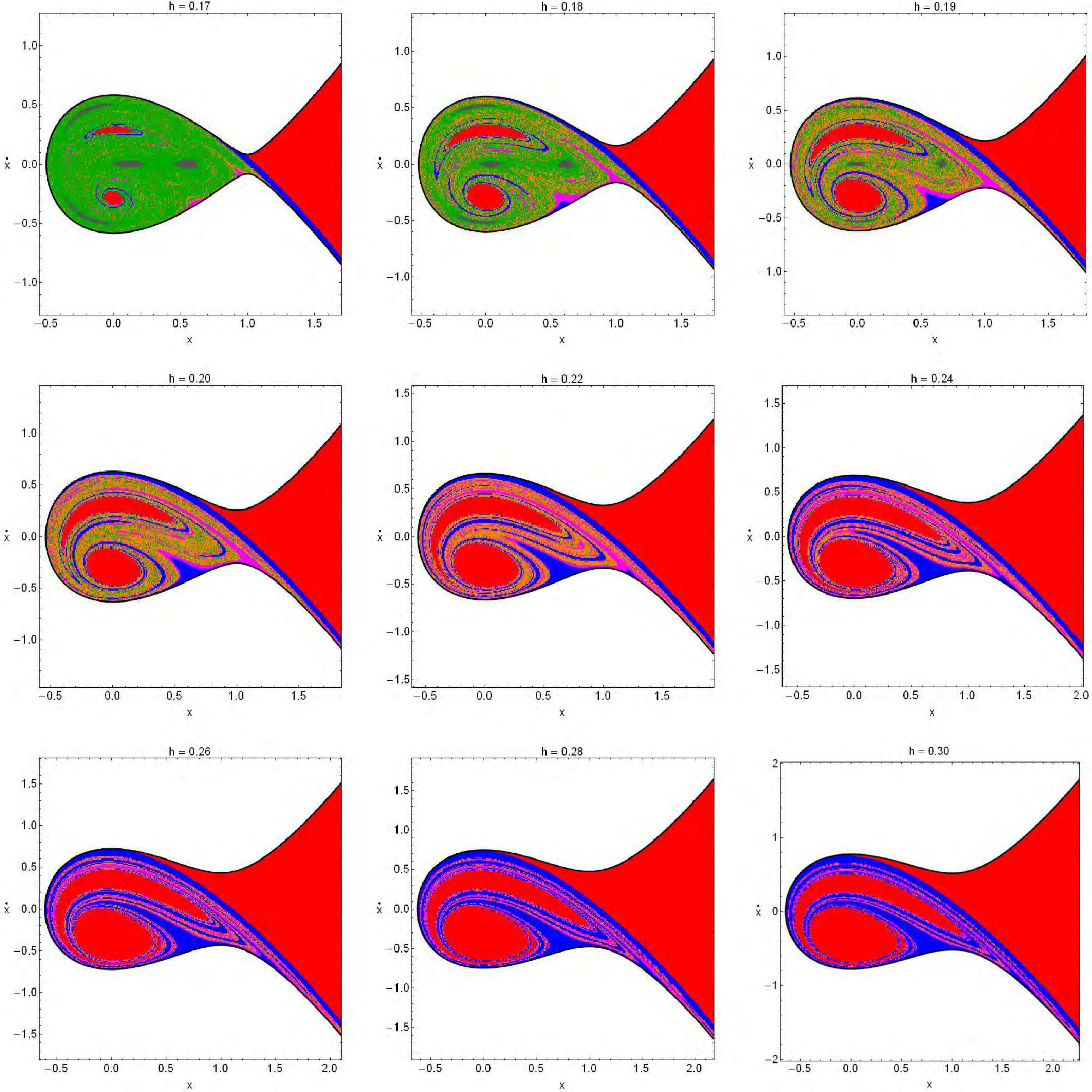}}
\caption{Color scale of the escape regions as a function of the number of intersections with the $y = 0$ axis upwards $(\dot{y} > 0)$. The color code is as follows: 0 intersections (red); 1 intersection (blue); 2 intersections (magenta); 3--10 intersections (orange); $> 10$ intersections (green). The gray regions represent stability islands of trapped orbits.}
\label{inter3}
\end{figure*}

\begin{figure}[!tH]
\includegraphics[width=\hsize]{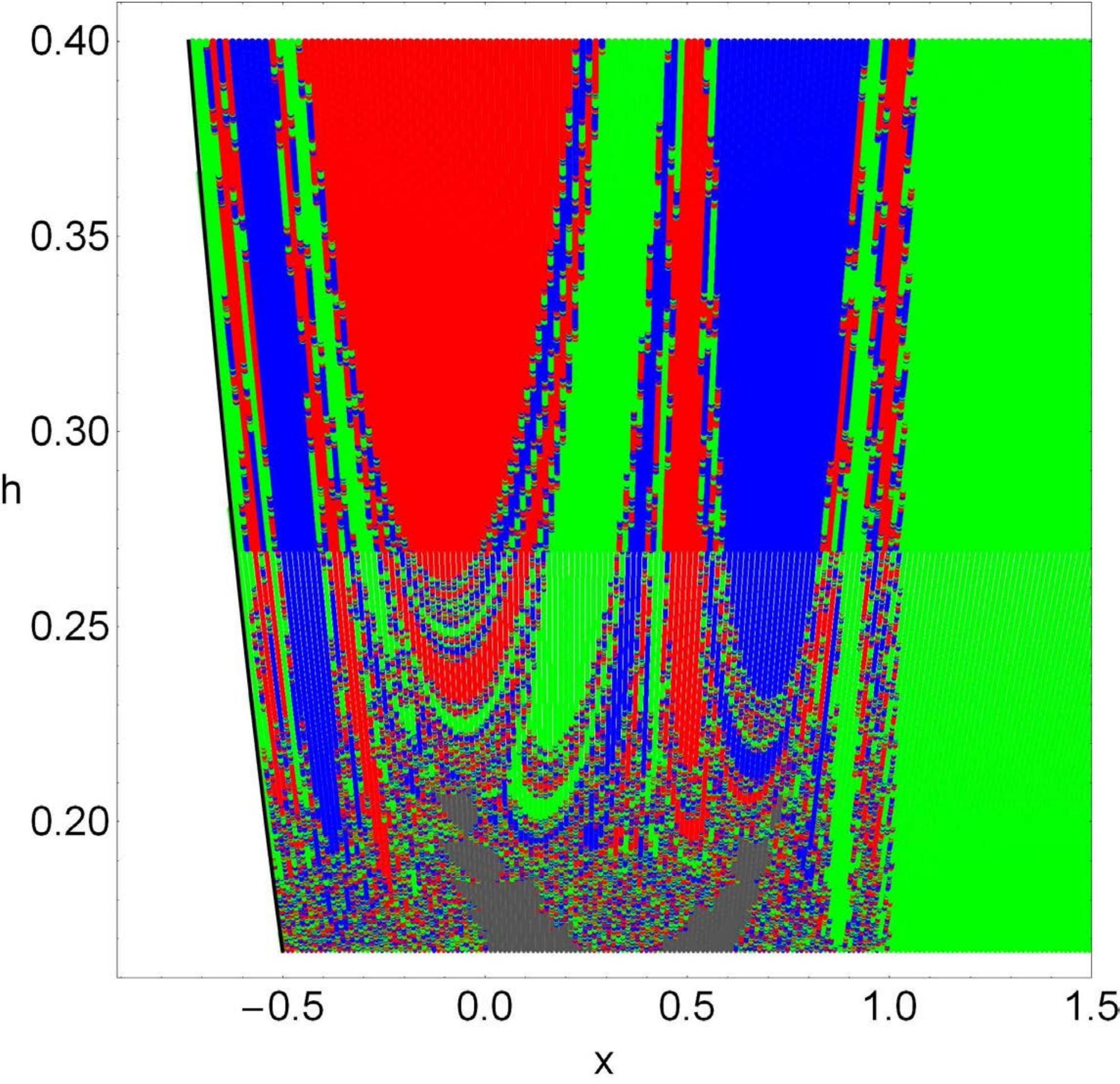}
\caption{Orbital structure of the $(x,h)$-plane when three channels of escape are present. This diagram gives a detailed analysis of the evolution of the trapped and escaping orbits of the dynamical system when the parameter $h$ changes. The color code is as in Fig. \ref{cxy3}.}
\label{xh3}
\end{figure}

In this case, we shall investigate the escape properties of unbounded motion of test particles for values of energy in the set $h = \{0.17, 0.18, 0.19, 0.20, 0.22, 0.24, 0.26, 0.28, 0.30\}$. We begin with initial conditions of orbits in the physical $(x,y)$ plane. The orbital structure of the physical plane for different values of the energy $h$ is show in Fig. \ref{cxy3}. Again, following the approach of the previous case, each initial condition is colored according to the escape channel through which the particular orbit escapes. Stability islands on the other hand, filled with initial conditions of orbits which do not escape are indicated as gray regions. We observe, that things are quite similar to that discussed previously in Fig. \ref{cxy2}. In fact, for energy levels very close to the escape energy, the central region of the plot is highly fractal and it is also occupied by several stability islands. However, as we increase the value of the energy the regions of regular trapped orbits are reduced, the physical plane becomes less and less fractal and well-defined basins of escape emerge.

The evolution of the percentages of trapped and escaping orbits on the physical $(x,y)$ plane when the value of the energy $h$ varies is presented in Fig. \ref{percs3}a. It is seen, that when $h = 0.17$, that is the first investigated energy level above the escape energy, escaping orbits through channels 2 and 3 share the same percentage (around 32\%), escapers through channel 1 have a slightly elevated percentage (around 34\%), while trapped orbits possess a very low rate corresponding only to 4\% of the physical plane. Once more, as we increase the value of the energy the rate of trapped orbits decreases and eventually vanishes for $h > 0.2$. Furthermore, we observe that the percentage of escaping orbits through channel 2 grows with increasing energy and for $h > 0.4$ it seems to saturate around 44\%. The percentage of escaping orbits through channel 3 on the other hand, exhibits a slow but constant decrease, while the rate of escaping orbits through exit 1 after small fluctuations it saturates around 32\% for $h > 0.4$. In general terms, we may conclude that throughout the energy range studied, the majority of orbits in the physical $(x,y)$ plane choose to escape through channel 2, while exit 3 seems to be the least favorable among the escape channels.

The following Fig. \ref{txy3} shows how the escape times $t_{\rm esc}$ of orbits are distributed on the $(x,y)$ plane. Light reddish colors correspond to fast escaping orbits, dark blue/purpe colors indicate large escape periods, while gray color denote trapped orbits. This grid representation of the physical plane gives us a much more clearer view of the orbital structure and especially about the trapped orbits. In particular, we see that for $h = 0.2$ we have the last indication of stability islands, as for all higher energy levels studied all orbits escape, thus defying basins of escape.

Our exploration continuous in the phase $(x,\dot{x})$ plane. The structure of the $(x,\dot{x})$ phase plane for different values of the energy is shown in Fig. \ref{cxpx3}. We observe a similar behavior to that discussed for the physical $(x,y)$ plane in Fig. \ref{cxy3}. Again, we can distinguish in the phase plane fractal regions where the prediction of the particular escape channel is impossible and regions occupied by escape basins. It is interesting to note, that the limiting curve (ZVC) is open at the right part due to the $x^3$ term entering the perturbation function. The rich fractal structure of the phase space shown in the grids of Fig. \ref{cxpx3} implies that our system has also a strong topological property, which is known as the Wada property. This special topological property has been identified and studied in several dynamical systems (e.g., [\citealp{AVS09}, \citealp{KY91}, \citealp{PCOG96}]) and it is a typical property in open Hamiltonian systems with three or more escape channels. An escape basin is a Wada basin if any boundary point also belongs to the boundary of at least two other basins [\citealp{BSBS12}, \citealp{KY91}]. It is seen in Fig. \ref{cxpx3} that for $h > 0.25$ all the KAM regime vanishes [\citealp{BBS08}] and therefore, all the initial conditions of orbits escape through one of the exits.

It is evident from Fig. \ref{percs3}b where the evolution of the percentages of trapped and escaping orbits on the phase plane as a function of the value of the energy $h$ is presented, that the pattern has many differences comparing to that discussed previously in Fig. \ref{percs3}a; only the percentage of trapped orbits exhibits similar behavior. To begin with, we observe that for $h = 0.17$ more than half of the phase plane (around 55\%) corresponds to initial conditions of orbits that escape through channel 1, while orbits escaping through exits 2 and 3 share about 44\% of the grid. As the value of the energy increases and we move away from the escape energy it is seen, that the rate of orbits escaping through exit 1 increases and always dominates, while on the hand, the percentages of orbits escaping through channels 2 and 3 drop. At the highest energy level studied $(h = 0.5)$, about 70\% of the total orbits escape through channel 1, about 20\% through channel 2 and only 10\% through channel 3. Thus, one may reasonably conclude that throughout the energy range studied, the vast majority of orbits in the phase $(x,\dot{x})$ plane choose to escape through channel 1, while channels 2 and 3 are much less likely to be chosen.

Fig. \ref{txpx3} shows the distribution of the escape times $t_{\rm esc}$ of orbits on the $(x,\dot{x})$ plane. It is evident, that orbits with initial conditions inside the exit basins escape from the system after short time intervals, or in other words, they possess extremely small escape periods. On the contrary, orbits with initial conditions located in the fractal parts of the phase plane need considerable amount of time in order to find one of the exits and escape. We see, that for $h > 0.2$ there is no indication of stability islands corresponding to trapped orbits. In another point of view, Fig. \ref{inter3} shows the regions of the phase plane which are now colored according to the number of intersections the orbits perform with the axis $y = 0$ upwards $(\dot{y} > 0)$. The red regions denote initial conditions of orbits that escape directly from the system without ever intersecting the $y = 0$ axis. The proportion of the total area on the phase plane occupied by these regions of direct escapes grows with increasing energy and for high enough energy levels they occupy more than 70\% of the grid. In Fig. \ref{xh3} we present the structure of the $(x,h)$-plane when $h \in (1/6, 1/2]$. It is seen, that trapped orbits exist only at low energies very close to the escape energy $(h < 0.22)$, while for larger energy levels all the orbits escape to infinity. Once more, highly fractal structure is observed near the stability islands of regular motion, while the degree of fractalization, or in other words the unpredictability of the system, reduces significantly where there are no trapped orbits and well-defined basins of escape cover the vast majority of the $(x,h)$-plane.

\begin{figure*}[!tH]
\centering
\resizebox{0.80\hsize}{!}{\includegraphics{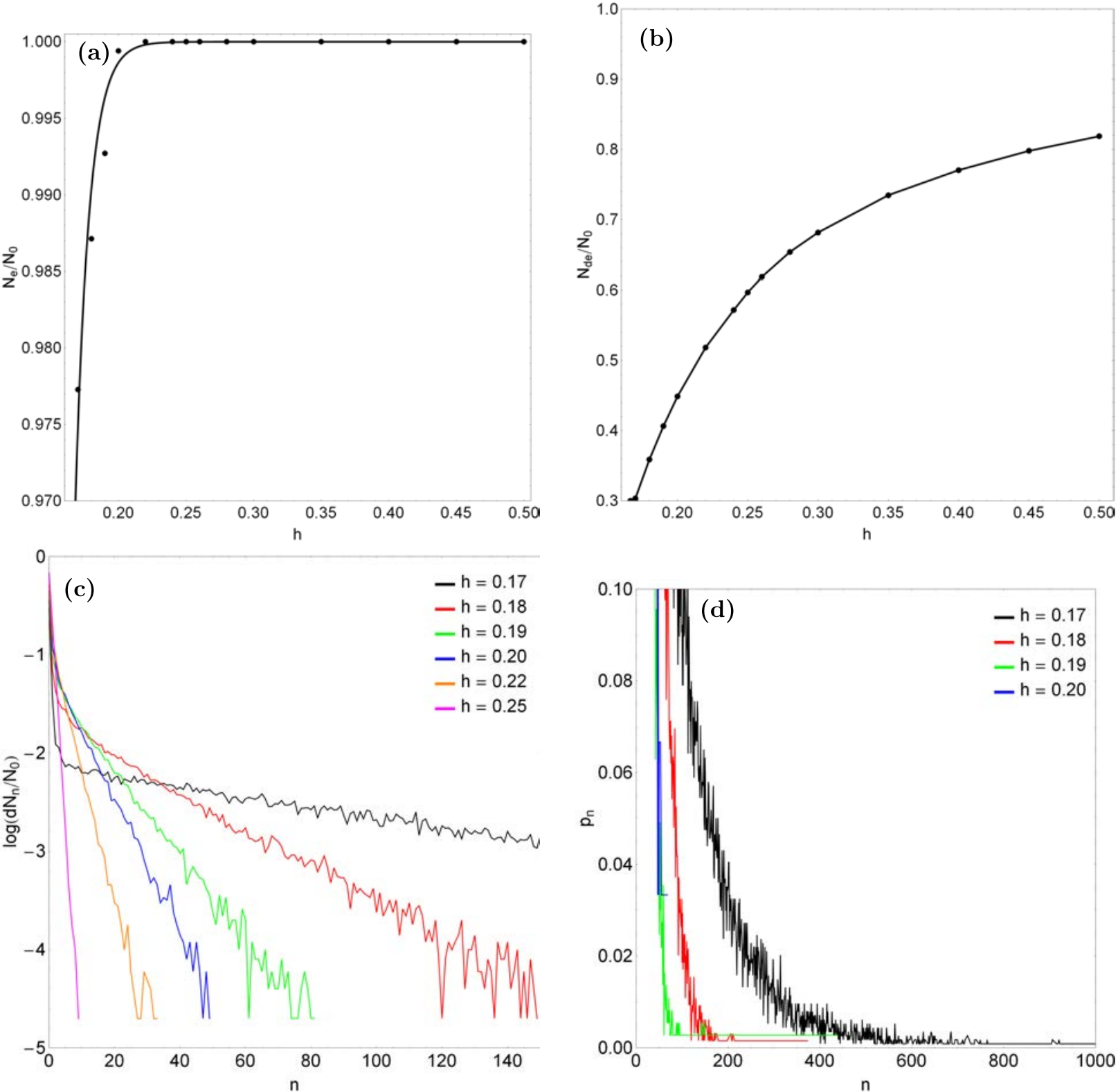}}
\caption{(a-upper left): Evolution of the proportion of escaping orbits $N_e/N_0$ as a function of the energy $h$, (b-upper right): Evolution of the proportion of directly escaping orbits $N_{de}/N_0$ as a function of the energy $h$, (c-lower left): Evolution of the logarithmic proportion $dN_n/N_0$ as a function of the number of the intersections $n$, for various values of the energy and (d-lower right): Evolution of the probability $p_n$ of escapes as a function of $n$ for several energy levels.}
\label{gm3}
\end{figure*}

At this point, we shall follow the approach discussed in subsection \ref{case2} in order to perform a statistical analysis of the escape process in the case of the $(x,\dot{x})$ phase plane for the Hamiltonian system with three channels of escape. Fig. \ref{gm3}a shows the proportion of escaping orbits $N_e/N_0$ as a function of the energy $h$. For values of energy beyond the escape energy, more than 95\% of the total orbits escape from the system. According to our numerical calculation, the evolution of the proportion of escaping orbits can be approximated by the formula
\begin{equation}
N_e/N_0(h) = 0.5\left[1 + \tanh\left(49h - 6.5\right)\right].
\label{tr31}
\end{equation}
In Fig. \ref{gm3}b we present the evolution of the direct escaping orbits $N_{de}/N_0$ as a function of the energy $h$. We see, that the amount of direct escaping orbits grows rapidly with increasing $h$ and for high energy levels $(h > 0.5)$ they populate about 80\% of the phase plane. The proportion of direct escapes can be given by the approximate formula
\begin{equation}
N_{de}/N_0(h) = -2.2 + 22.86 h - 59.24 h^2 + 51.42 h^3.
\label{tr32}
\end{equation}
Furthermore, Fig. \ref{gm3}c depicts the logarithm of the proportion of escaping orbits $dN_n/N_0$, as a function of the intersections with the $y = 0$ axis upwards an orbit performs before it escapes. We observe, that the escape time of orbits decreases with increasing $n$. In particular, the escape rates are high for relatively small $n$, while they drop rapidly for larger $n$. Finally, we calculated the probability of escape as a function of the number of intersections for various values of the energy. The evolution of $p_n$ as a function of $n$ for various energy levels is shown in Fig. \ref{gm3}d.

\subsection{Case III: Four channels of escape}
\label{case3}

\begin{figure*}[!tH]
\centering
\resizebox{\hsize}{!}{\includegraphics{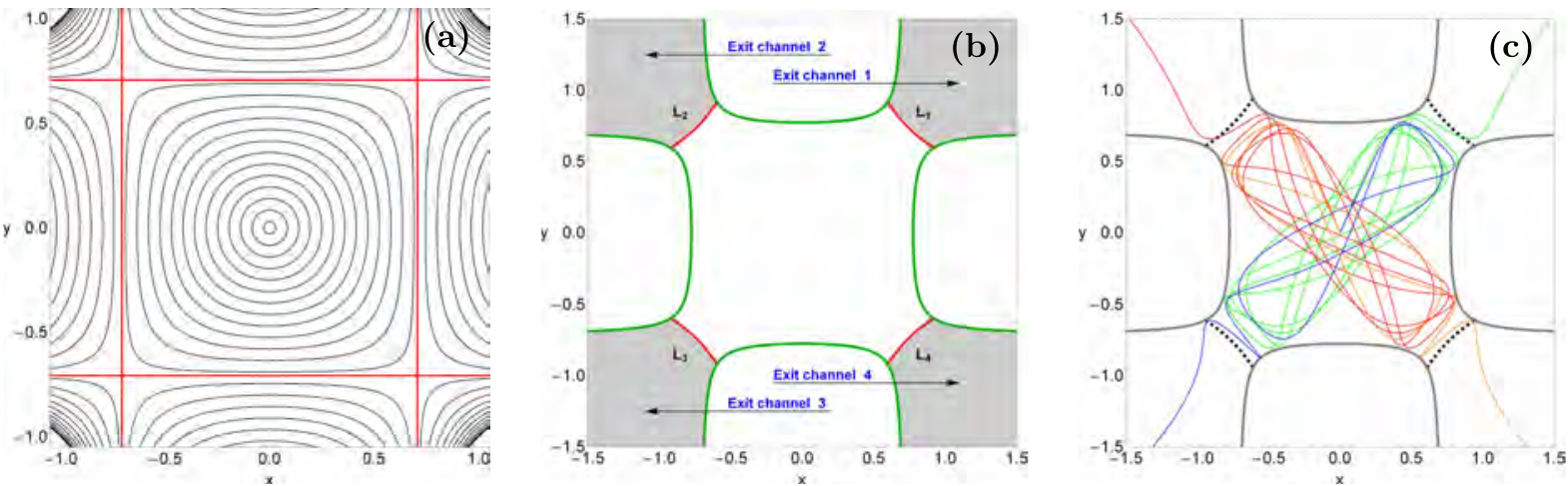}}
\caption{(a): Equipotential curves of the potential (\ref{pot}) for various values of the energy $h$ when $V_1(x,y) = - x^2 y^2$. The equipotential curve corresponding to the energy of escape is shown with red color; (b): The open ZVC at the physical $(x,y)$ plane when $h = 0.3$. $L_1$, $L_2$, $L_3$ and $L_4$ indicate the four unstable Lyapunov orbits plotted in red; (c): Four escaping orbits when $h = 0.3$. The orbit which escapes from channel 1 is potted with green color, the orbit escaping from channel 2 with red color, the one from channel 3 with blue, while orange color is used for the orbits which escapes through channel 4.}
\label{exit4}
\end{figure*}

The last case under investigation is a Hamiltonian system with four channels of escape. In order to obtain this number of exits in the limiting curve in the physical $(x,y)$ plane, we chose the perturbation term $V_1(x,y) = - x^2 y^2$ and the corresponding Hamiltonian is he following
\begin{equation}
H_3 = \frac{1}{2}\left(\dot{x}^2 + \dot{y}^2 + x^2 + y^2\right) - x^2 y^2 = h.
\label{ham3}
\end{equation}
The Hamiltonian $H_3$ is invariant under $x \rightarrow - x$ and/or $y \rightarrow - y$. The escape mechanism in this particular Hamiltonian system with the four escape channels and escape energy equals to 1/4 has already been examined (e.g., [\citealp{C90}, \citealp{CK92}, \citealp{CKK93}, \citealp{KSCD99}, \citealp{NH01}]). In Fig. \ref{exit4}a, we present the equipotential curves of the potential (\ref{pot}) for various values of the energy $h$, while the equipotential corresponding to the energy of escape $h_{esc}$ is plotted with red color in the same plot. In addition, the open ZVC at the physical $(x,y)$ plane when $h = 0.3 > h_{esc}$ is given with green color in Fig. \ref{exit4}b, while the four channels of escape are also shown. In the same figure, the four unstable Lyapunov orbits $L_1$, $L_2$, $L_3$ and $L_4$ are denoted using red color. In Fig. \ref{exit4}c we plotted with different colors four orbits, one escaping from channel 1, one from channel 2, one from channel 3 and the last one from channel 4, when $h = 0.3$.

\begin{figure*}[!tH]
\centering
\resizebox{0.90\hsize}{!}{\includegraphics{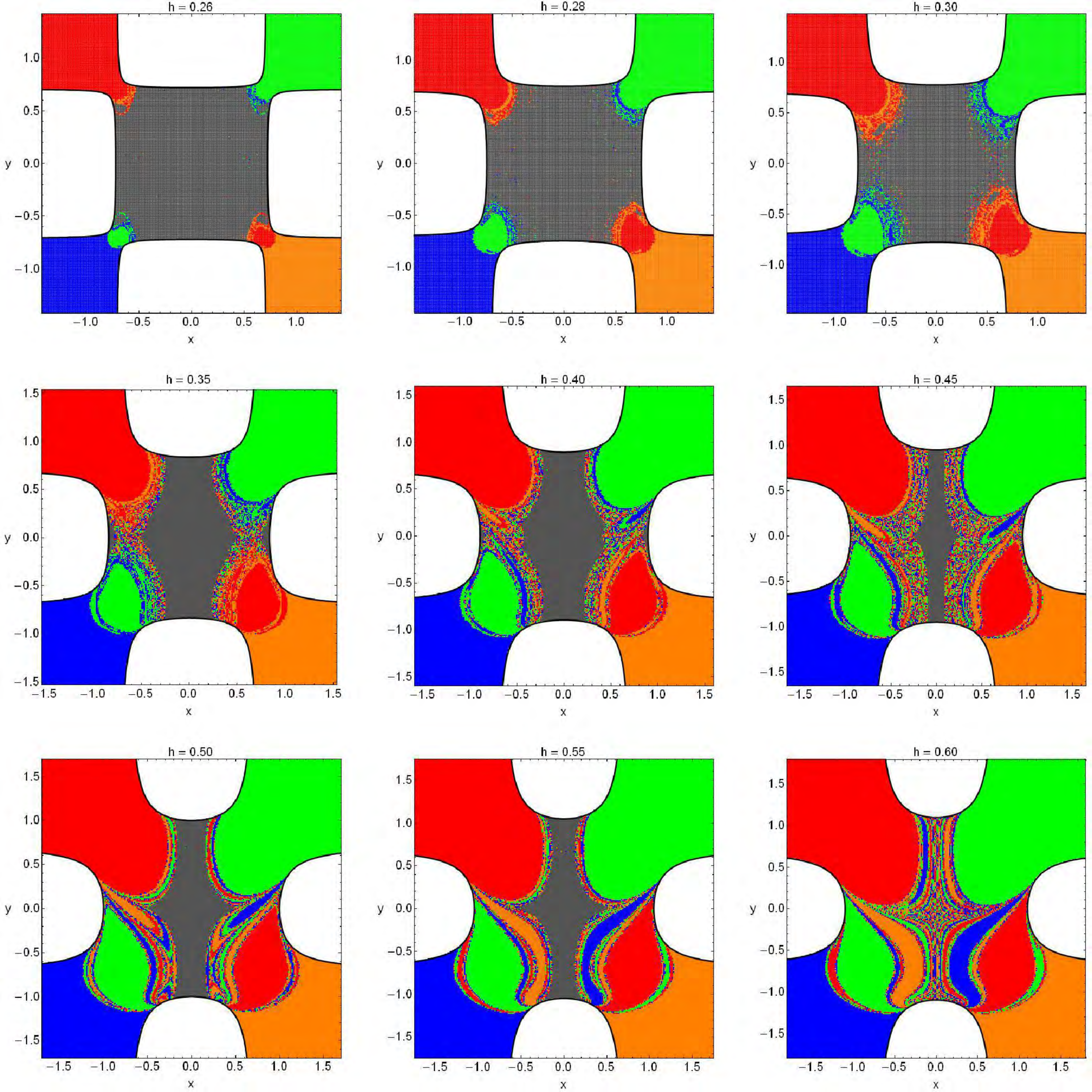}}
\caption{The structure of the physical $(x,y)$ plane for several values of the energy $h$, distinguishing between different escape channels. The color code is as follows: Trapped (gray); escape through channel 1 (green); escape through channel 2 (red); escape through channel 3 (blue); escape through channel 4 (orange).}
\label{cxy4}
\end{figure*}

\begin{figure*}[!tH]
\centering
\resizebox{0.8\hsize}{!}{\includegraphics{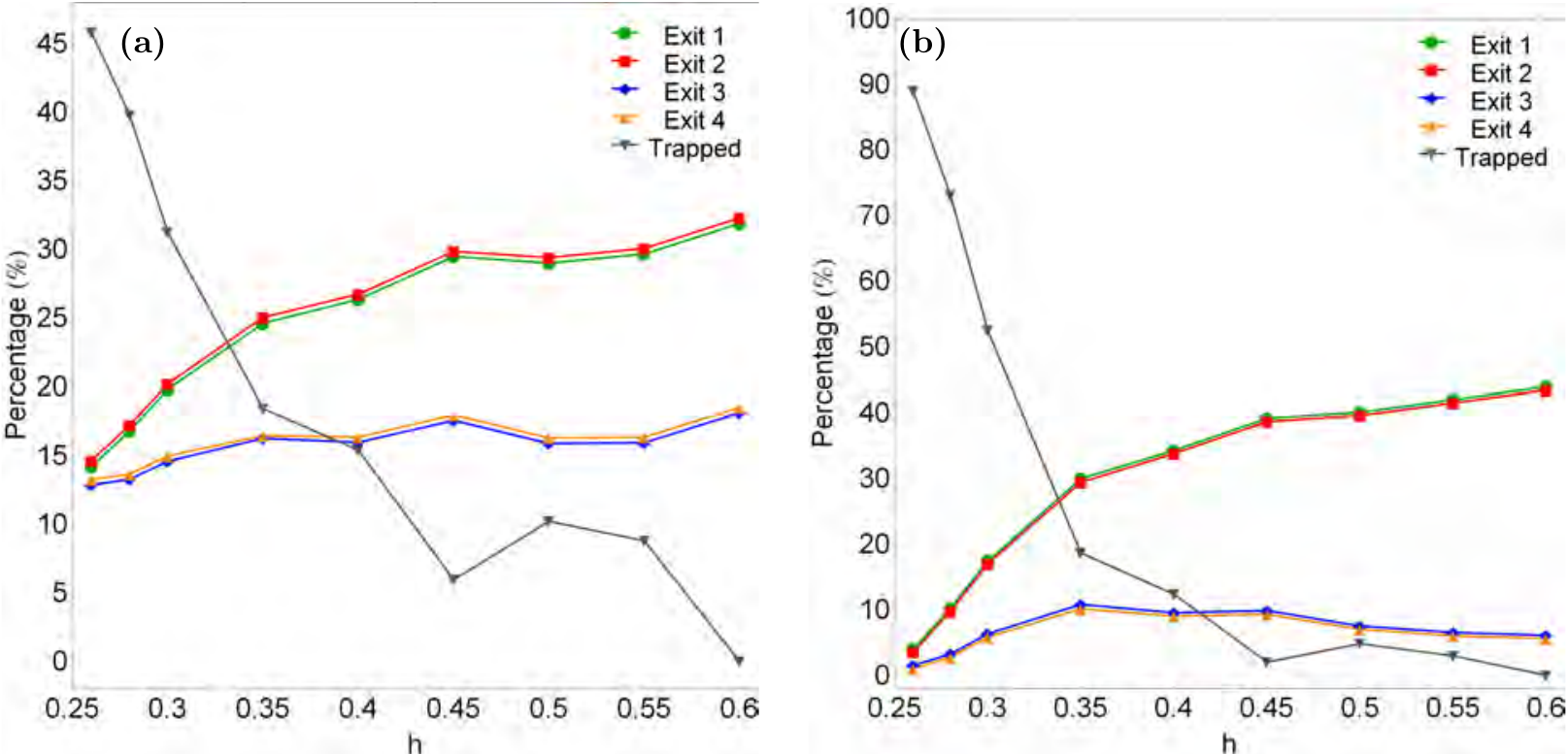}}
\caption{Evolution of the percentages of trapped and escaping orbits when varying the energy $h$ (a-left): on the physical $(x,y)$ plane and (b-right): on the phase $(x,\dot{x})$ plane.}
\label{percs4}
\end{figure*}

\begin{figure*}[!tH]
\centering
\resizebox{0.95\hsize}{!}{\includegraphics{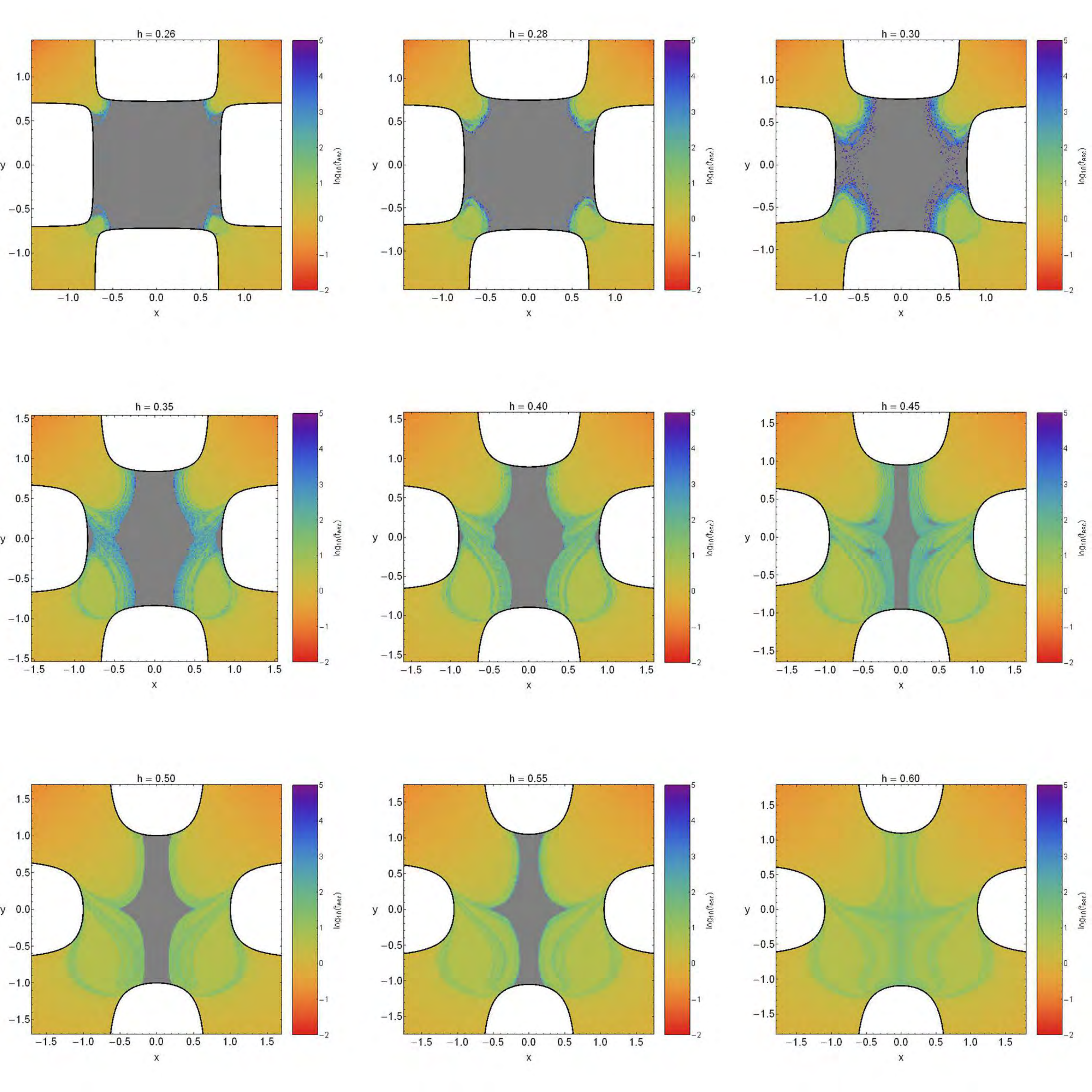}}
\caption{Distribution of the escape times $t_{\rm esc}$ of the orbits on the $(x,y)$ plane. The darker the color, the larger the escape time. Trapped orbits are indicated by gray color.}
\label{txy4}
\end{figure*}

\begin{figure*}[!tH]
\centering
\resizebox{0.90\hsize}{!}{\includegraphics{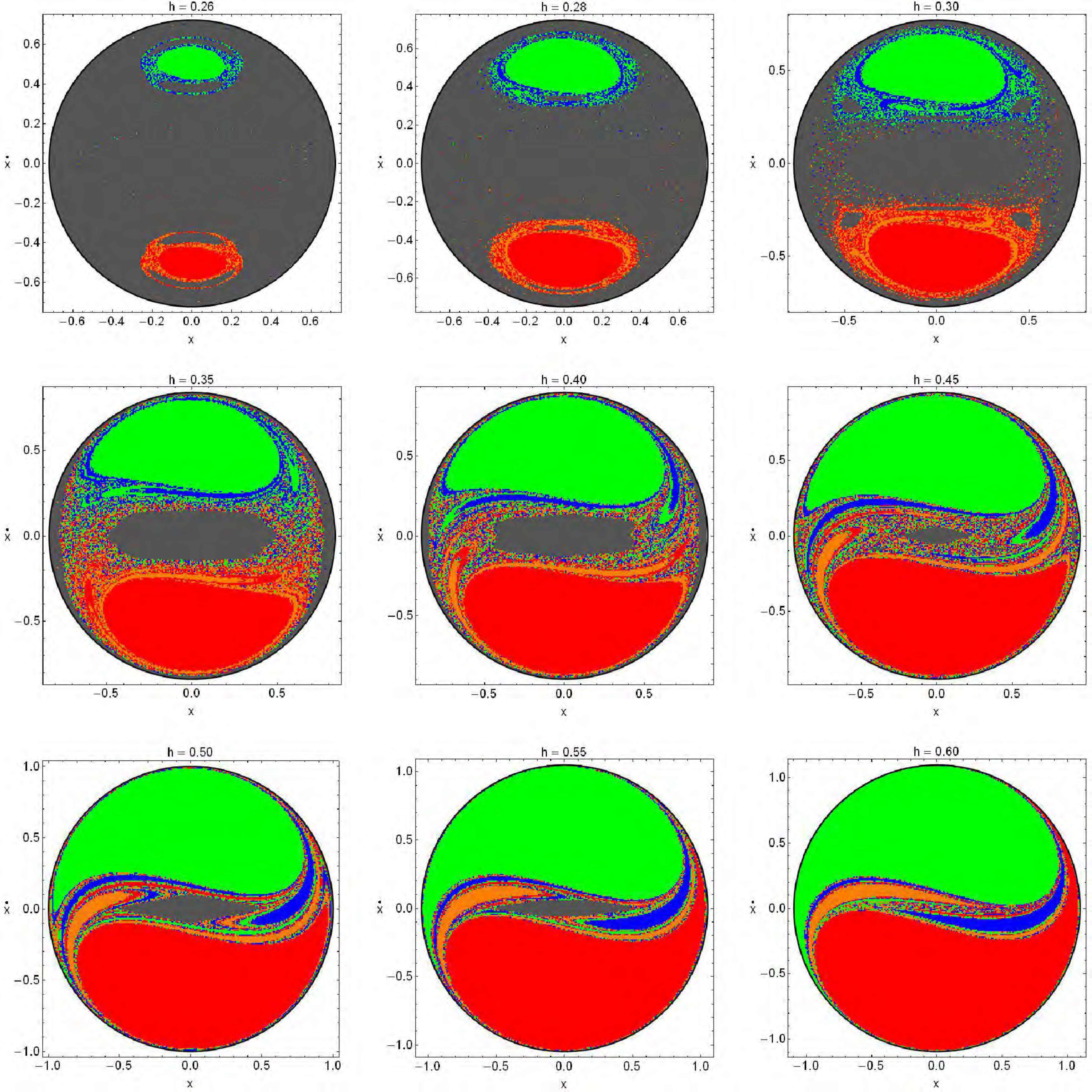}}
\caption{The structure of the phase $(x,\dot{x})$ plane for several values of the energy $h$, distinguishing between different escape channels. The color code is as follows: Trapped (gray); escape through channel 1 (green); escape through channel 2 (red); escape through channel 3 (blue); escape through channel 4 (orange).}
\label{cxpx4}
\end{figure*}

\begin{figure*}[!tH]
\centering
\resizebox{0.80\hsize}{!}{\includegraphics{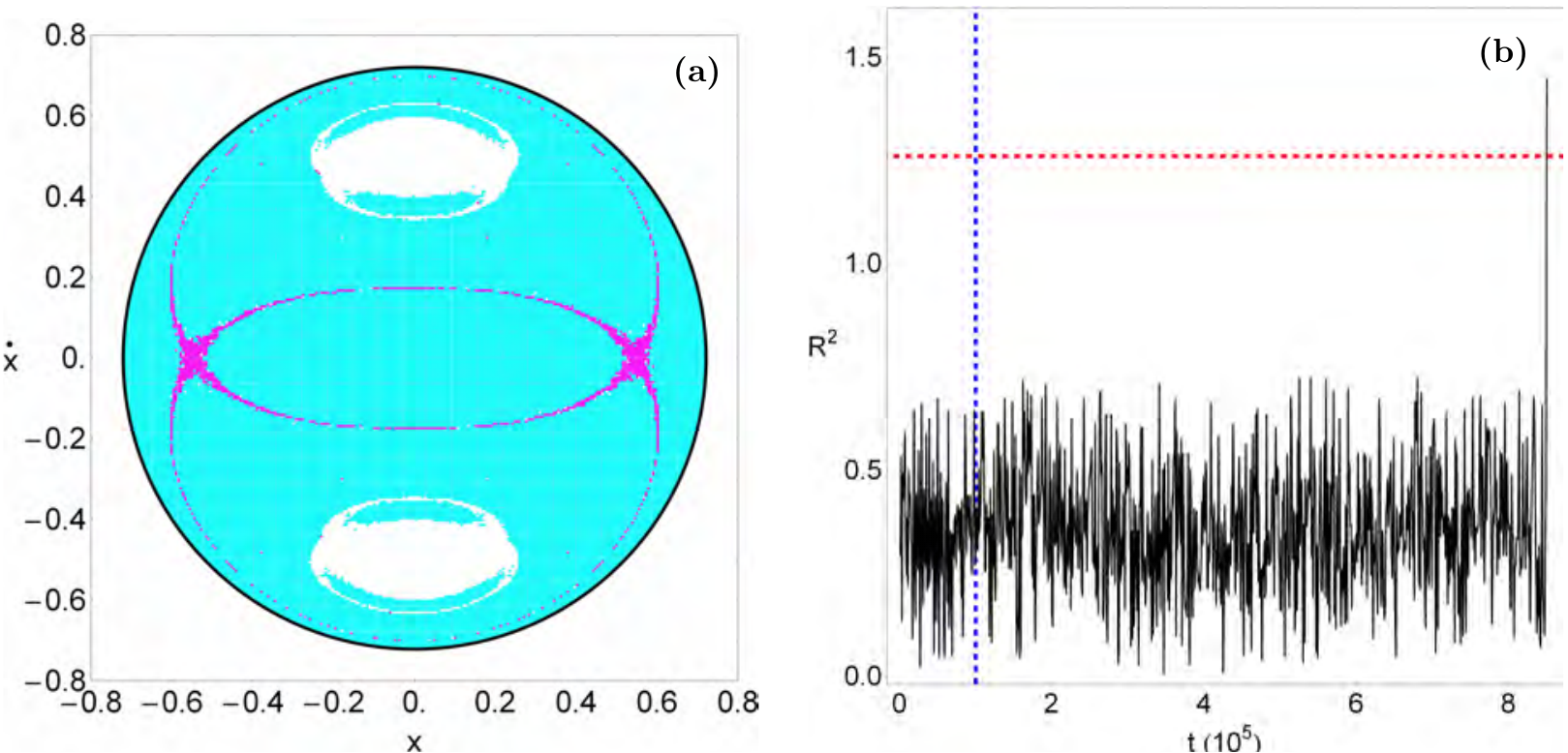}}
\caption{(a-left): The structure of the phase $(x,\dot{x})$ plane for $h = 0.26$, distinguishing between trapped regular orbits (cyan), trapped chaotic orbits (magenta) and escaping orbits (white). (b-left): Time evolution of $R^2 = x^2 + y^2$ for a super sticky orbit when $h = 0.26$. The horizontal, red, dashed line at 1.26 approximates the position of the unstable Lyapunov orbits at the four exits, while the vertical, blue, dashed line denotes the initial integration interval of $10^5$ time units.}
\label{RC}
\end{figure*}

\begin{figure*}[!tH]
\centering
\resizebox{0.95\hsize}{!}{\includegraphics{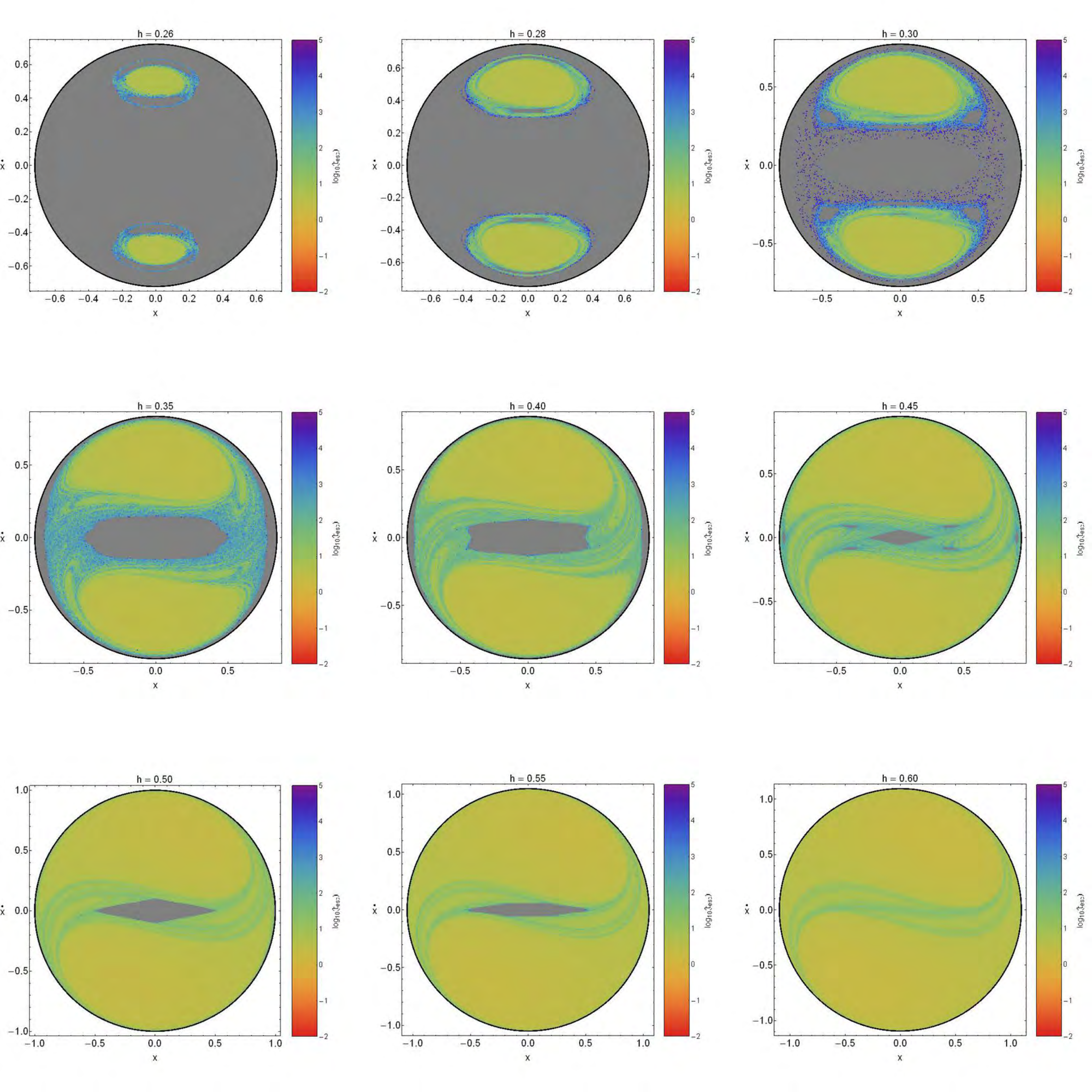}}
\caption{Distribution of the escape times $t_{\rm esc}$ of the orbits on the $(x,\dot{x})$ plane. The darker the color, the larger the escape time. Trapped orbits are indicated by gray color.}
\label{txpx4}
\end{figure*}

\begin{figure*}[!tH]
\centering
\resizebox{0.90\hsize}{!}{\includegraphics{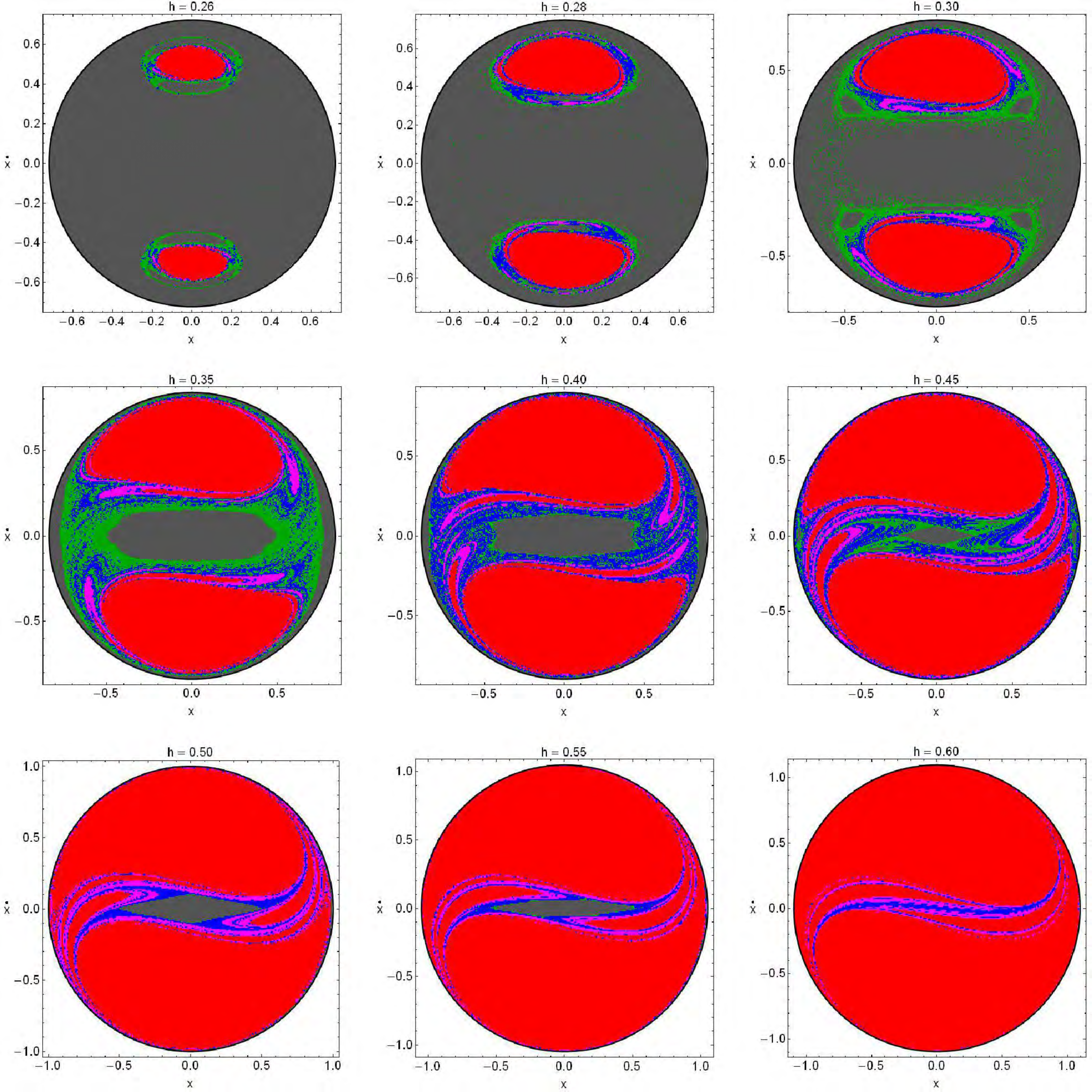}}
\caption{Color scale of the escape regions as a function of the number of intersections with the $y = 0$ axis upwards $(\dot{y} > 0)$. The color code is as follows: 0 intersections (red); 1 intersection (blue); 2 intersections (magenta); 3--10 intersections (orange); $> 10$ intersections (green). The gray regions represent stability islands of trapped orbits.}
\label{inter4}
\end{figure*}

\begin{figure}[!tH]
\includegraphics[width=\hsize]{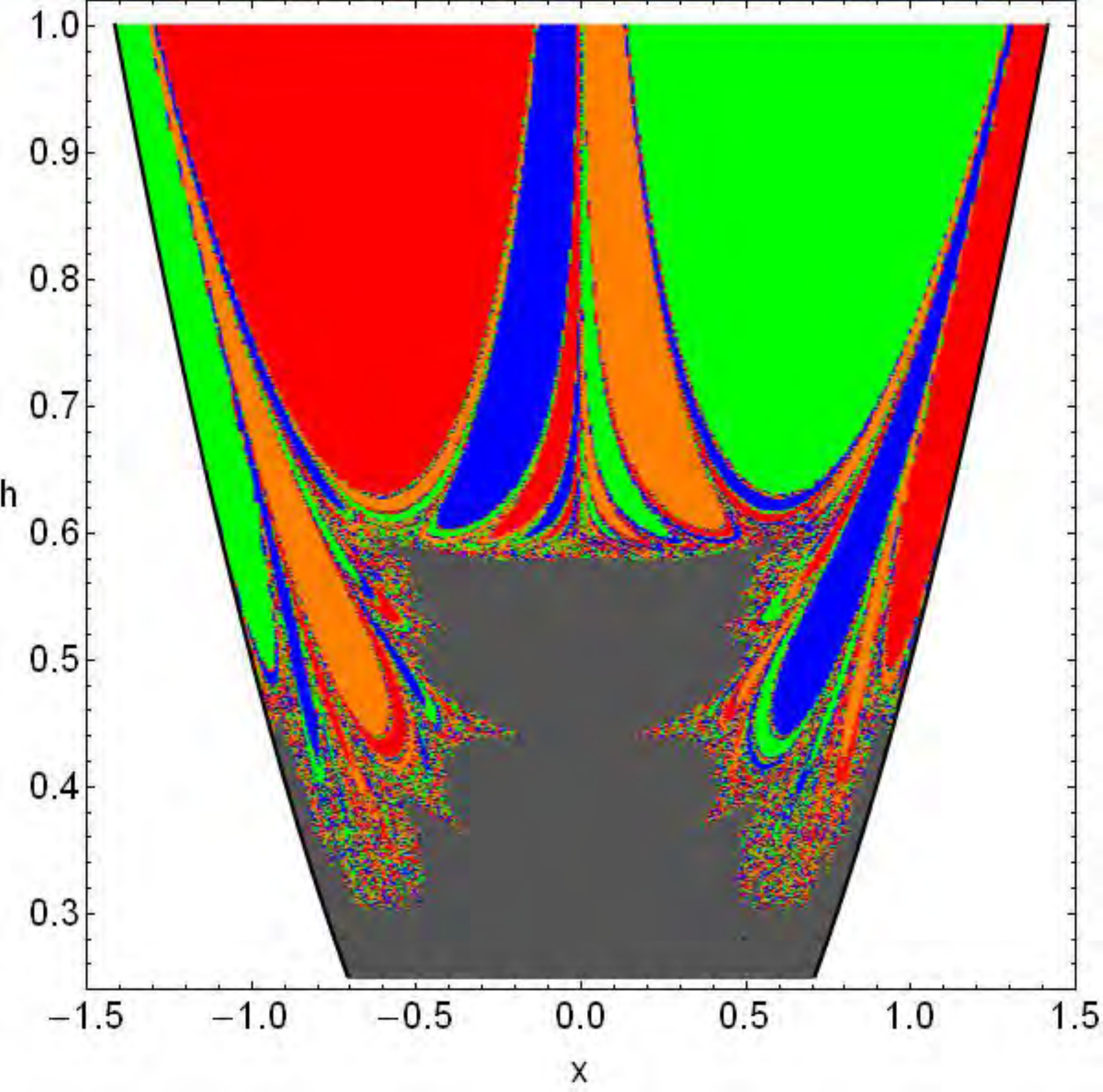}
\caption{Orbital structure of the $(x,h)$-plane when four escape channels are present. This diagram gives a detailed analysis of the evolution of the trapped and escaping orbits of the dynamical system when the parameter $h$ changes. The color code is as in Fig. \ref{cxy4}.}
\label{xh4}
\end{figure}

The escape properties and mechanism of unbounded motion of test particles for values of energy in the set $h = \{0.26, 0.28, 0.30, 0.35, 0.40, 0.45, 0.50, 0.55, 0.60\}$ will be examined. We begin, as usual, with initial conditions of orbits in the physical $(x,y)$ plane. Fig. \ref{cxy4} shows the orbital structure of the physical plane for different values of the energy $h$. Again, following the same approach of the previous cases, each initial condition is colored according to the escape channel through which the particular orbit escapes. Areas corresponding to trapped orbits on the other hand, are indicated as gray regions. It is evident, that the structure of the $(x,y)$ plane differs significantly with respect to the plots shown previously in Figs. \ref{cxy2} and \ref{cxy3}. We see, that for values of energy very close to the escape energy almost all the central region of the grid is covered by initial conditions of trapped orbits, while escaping orbits exist only near the four exits. However, with increasing energy the area on the physical plane occupied by trapped orbits reduces and several basins of escape begin to emerge. At the highest energy level studied $(h = 0.6)$, there is no indication of trapped motion and all orbits escape to infinity through one of the four escape channels. We also observe, the existence of well-formed basins of escape, while the central region of the grid still remains highly fractal. Here we should like to note, that in general terms, throughout the energy range the structure of the physical plane $(x,y)$ is symmetrical with respect to the $x = 0$ axis.

It is of particular interest to monitor the evolution of the percentages of trapped and escaping orbits on the physical $(x,y)$ plane when the value of the energy $h$ varies. A diagram depicting this evolution is presented in Fig. \ref{percs4}a. We see, that for $h = 0.26$, that is an energy level just above the escape energy, about 50\% of the physical plane is covered by initial conditions of trapped orbits. As the value of the energy increases however, the rate of trapped orbits drops rapidly and eventually at $h = 0.6$ it vanishes. We also observe, that the evolution of the percentages of orbits escaping through channels 1 and 3 coincide with the evolution of the percentages escaping through channels 2 and 4, respectively. We anticipated this behaviour of the escape percentages, which is a natural result of the symmetrical structure of the $(x,y)$ plane. It is seen, that initially $(h = 0.26)$ all rates of escaping orbits coincide at about 14\%. Then, with increasing energy the rates of escaping orbits increase and also start to diverge. At the highest energy studied, escaping orbits through channels 1 and 2 share about 65\% of the physical plane, while escaping orbits trough channels 3 and 4 occupy the remaining 35\% of the grid. Therefore, one may reasonably conclude that in general terms, throughout the range of the values of the energy studied, the majority of orbits in the physical $(x,y)$ plane choose to escape either through channel 1 or channel 2.

The distribution of the escape times $t_{\rm esc}$ of orbits on the physical plane is given in Fig. \ref{txy4}. Light reddish colors correspond to fast escaping orbits, dark blue/purpe colors indicate large escape periods, while gray color denote trapped orbits. Here, we have a better view regarding the amount of trapped orbits. Indeed, we see that for $h = 0.6$ all orbits escape from the system. Moreover we observe, that orbits with initial conditions close to the area occupied by trapped orbits have significantly large escape periods, while on the other hand, orbits located near the escape channels escape very quickly having escaping rates of about two orders smaller.

We continue our investigation to the phase $(x,\dot{x})$ plane, the structure of which for different values of the energy is presented in Fig. \ref{cxpx4}. One may observe, that for $h < 0.3$ most of the phase plane is covered by a vast region corresponding to trapped orbits, while only two small islands of initial conditions of escaping orbits exit. However, as the value of the energy increases and we move far away for the escape energy, the extent of these two islands grows and for $h > 0.35$ the trapped orbits are mainly confined to the central region of the phase plane. At the same time, small elongated spiral basins of escape emerge inside the fractal region which surrounds the area of trapped orbits. Furthermore, at very high energy levels $(h > 0.55)$ we see that trapped orbits disappear completely from the grid and the two main basins of escape take over the vast majority of the phase plane, while the elongated escape basins remain confined to the central region. As we noticed previously when discussing the physical $(x,y)$ plane, there is also a symmetry in the phase plane. In particular, throughout the energy range the structure of the phase plane $(x,\dot{x})$ is somehow symmetrical (not with the strick sense) with respect to the $\dot{x} = 0$ axis.

The evolution of the percentages of trapped and escaping orbits on the phase plane as a function of the value of the energy $h$ is given in Fig. \ref{percs4}b. For $h = 0.26$, we see that trapped orbits dominate the phase plane as they occupy about 90\% of the gird. However as usual, with increasing energy the dominance of trapped orbits deteriorates rapidly due to the increase of the rates of escaping orbits which form basins of escape. We observe, that once more as in Fig. \ref{percs4}a, the evolution of the percentages of orbits escaping through channels 1 and 3 coincides with the evolution of the percentages escaping trough channels 2 and 4, respectively. The percentages of all types of escaping orbits increase but with different rates and for $h > 0.35$ they overwhelm the amount of trapped orbits. In particular, we see that the percentages of orbits escaping through exits 1 and 2 are always higher than those corresponding to orbits escaping through channels 3 and 4. Moreover, the rates of exits 1 and 2 increase constantly and at the highest energy level studied $(h = 0.6)$ the share about 90\% of the entire phase plane. On the other hand, the percentages of exits 3 and 4, even though they also grow with increasing energy, they always possess significantly smaller values than exits 1 and 2 and for $h > 0.4$ they seem to saturate around 5\%. Thus, we may conclude that the vast majority of orbits in the phase $(x,\dot{x})$ plane exhibit clear sings of preference through exits 1 and 2, while channels 3 and 4 have considerable less probability to be chosen.

Taking into account that for low values of the energy there is a considerable amount of trapped motion in the phase plane, we decided to use the SALI method in order to distinguish between regular and chaotic trapped orbits. In Fig. \ref{RC}a we present the phase plane for $h = 0.26$, where each initial condition is plotted according to the regular (cyan) or chaotic (magenta) character of the orbit, while white areas correspond to escaping orbits. It is seen, that the vast majority of the trapped orbits are regular however, a thin layer composed of chaotic trapped orbits is also present. Therefore, a natural and very important question arises: do these chaotic bounded orbits remain trapped forever? Remember, that in the current investigation we set the maximum time of the numerical integration to be equal to $10^5$ time units. We suspect, that all these trapped chaotic orbits will eventually escape from the system if they have enough time to evolve. Thus, in order to shed some light to this issue, we let the time running and we integrated these orbits until they escape. Our numerical calculations revealed, that these orbits are in fact super sticky orbits which possess extremely high escape periods up to $3.5 \times 10^6$ time units. A characteristic example of such a super sticky orbit is given in Fig. \ref{RC}b, where we monitor the time evolution of $R^2 = x^2 + y^2$. The horizontal, red, dashed line at 1.26 approximates the position of the unstable Lyapunov orbits at the four exits, while the vertical, blue, dashed line denotes the initial integration time $(10^5$ time units). We see, that the particular orbit escapes through channel 2 after a time interval of about 851000 time units which is more than 8.5 times the initial integration period. Our additional computations indicate, that these super sticky orbits correspond to less than 10\% of the total (regular plus chaotic) trapped orbits so, using $10^5$ time units for the numerical integration and counting them as trapped, even though they escape after vast time intervals, does not have a huge impact in our results.

The following Fig. \ref{txpx4} shows the distribution of the escape times $t_{\rm esc}$ of orbits on the $(x,\dot{x})$ plane. It is clear, that orbits with initial conditions inside the exit basins escape to infinity after short time intervals, or in other words, they possess extremely small escape periods. On the contrary, orbits with initial conditions located in the fractal parts of the phase plane need considerable amount of time in order to find one of the four exits and escape. It is seen, that at the highest energy level studied $(h = 0.6)$ there is no indication of bounded motion and all orbits escape to infinity sooner or later.

In Fig. \ref{inter4}, we reconstructed the grids in the phase plane using different color code and now the regions of the phase plane are colored according to the number of intersections the orbits perform with the axis $y = 0$ upwards $(\dot{y} > 0)$. Specifically, red regions correspond to initial conditions of orbits that escape directly from the system without any intersection with the $y = 0$ axis. We see, that the proportion of the total area on the phase plane occupied by orbits which escape directly from the system grows rapidly with increasing energy and for $h > 0.6$ they occupy more than 90\% of the entire grid. Fig. \ref{xh4} depicts the structure of the $(x,h)$-plane when $h \in (0.25, 1]$. At low energy levels, one may observe, three important issues: (i) the vast majority of oribis are trapped, (ii) the structure of the $(x,h)$-plane exhibits a high degree of fractalization and (iii) basins of escape are present only at the outer parts of the grid. However, when $h > 0.6$, that is when trapped orbits cease to exist we see that the fractal structure disappears and all the $(x,h)$-plane is coved by well-defined basins of escape. It should also be pointed out, that the structure of the $(x,h)$-plane is symmetrical with respect to the $x = 0$ axis.

\begin{figure*}[!tH]
\centering
\resizebox{0.80\hsize}{!}{\includegraphics{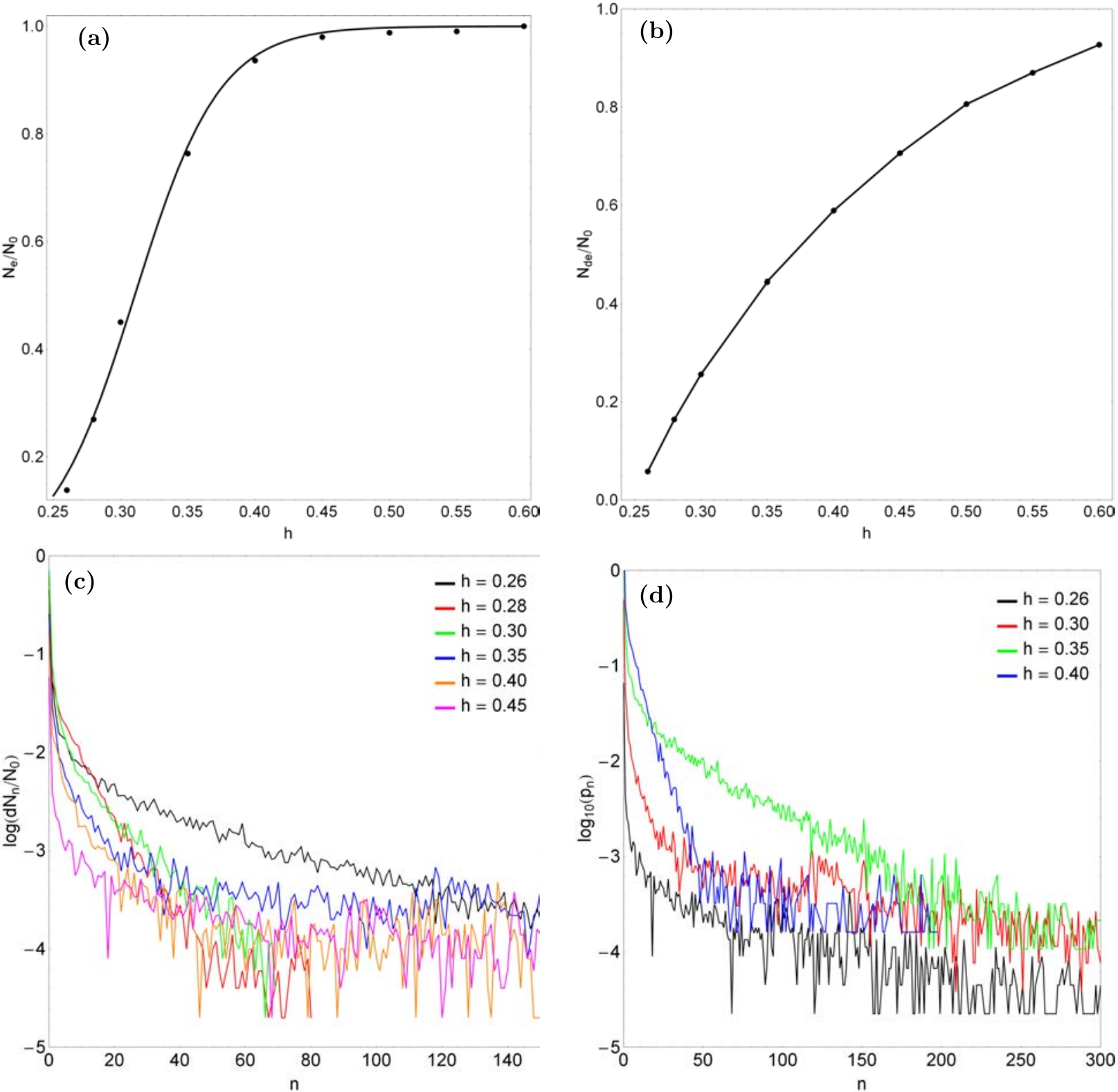}}
\caption{(a-upper left): Evolution of the proportion of escaping orbits $N_e/N_0$ as a function of the energy $h$, (b-upper right): Evolution of the proportion of directly escaping orbits $N_{de}/N_0$ as a function of the energy $h$, (c-lower left): Evolution of the logarithmic proportion $dN_n/N_0$ as a function of the number of the intersections $n$, for various values of the energy and (d-lower right): Evolution of the probability $p_n$ of escapes as a function of $n$ for several energy levels.}
\label{gm4}
\end{figure*}

Before closing this section, we would like to perform a statistical analysis of the escape process in the case of the $(x,\dot{x})$ phase plane for the Hamiltonian system with four channels of escape. The proportion of escaping orbits $N_e/N_0$ as a function of the energy $h$ is presented in Fig. \ref{gm4}a. We see that for $h > 0.45$, more than 90\% of the total orbits escape from the system. Our numerical computations suggest, that the evolution of the proportion of escaping orbits can be approximated by the formula
\begin{equation}
N_e/N_0(h) = 0.5\left[1 + \tanh\left(15.85h - 4.93\right)\right].
\label{tr41}
\end{equation}
Furthermore, in Fig. \ref{gm4}b we present the evolution of the direct escaping orbits $N_{de}/N_0$ as a function of the energy $h$. As it was found in the previously examined cases, the amount of direct escaping orbits grows rapidly with increasing $h$ and for high energy levels $(h > 0.6)$ they populate more than 90\% of the phase plane. The proportion of direct escapes can be given by the approximate formula
\begin{equation}
N_{de}/N_0(h) = -1.496 + 7.585 h - 5.938 h^2.
\label{tr42}
\end{equation}
The evolution of the logarithm of the proportion of escaping orbits $dN_n/N_0$, as a function of the intersections with the $y = 0$ axis upwards an orbit performs before it escapes is given in Fig. \ref{gm4}c. One may observe, that the escape time of orbits decreases with increasing $n$. Being more precise, the escape rates are high enough for relatively small number of intersections $n$, while they fall rapidly for larger $n$. Finally, we computed the probability of escape as a function of the number of intersections for various values of the energy $h$. Our results are shown in Fig. \ref{gm4}d, where we present the evolution of $p_n$ as a function of $n$ for various energy levels.

\section{Conclusions and discussion}
\label{disc}

The main objective of this work was to review but also numerically investigate even further the escape properties of orbits in a dynamical system of two-dimensional coupled perturbed harmonic oscillators, which is a characteristic example of open Hamiltonian systems. The key feature of this type of Hamiltonians is that they have a finite energy of escape. In particular, for energies smaller than the escape value, the equipotential surfaces are close and therefore escape is impossible. For energy levels larger than the escape energy however, the equipotential surfaces open and several channels of escape appear through which the particles can escape to infinity. Here we should emphasize, that if a test particle has energy larger than the escape value, this does not necessarily mean that the particle will certainly escape from the system and even if escape does occur, the time required for an orbit to cross a Lyapunov orbit and hence escape to infinity may be vary long compared with the natural crossing time. The function containing the perturbation terms affects significantly the structure of the equipotential curves and determines the exact number of the escape channels. We chose such forms of perturbations and divided our study into three cases with respect to the number of the escape channels.

Since a distribution function of the system was not available so as to use it for extracting the different samples of orbits, we had to follow an alternative path. We defined for each set of values of the energy, dense grids of initial conditions regularly distributed in the area allowed by the value of the energy in both the physical and the phase space. In both cases, the density of the grids was controlled in such a way that always there are about 50000 orbits to be examined. For the numerical integration of the orbits in each grid, we needed roughly between 1 minute and 3 days of CPU time on a Pentium Dual-Core 2.2 GHz PC, depending both on the amount of trapped orbits and on the escape rates of orbits in each case. For each initial condition, the maximum time of the numerical integration was set to be equal to $10^5$ time units however, when a particle escapes the numerical integration is effectively ended and proceeds to the next initial condition.

The structure of both the physical $(x,y)$ and phase $(x,\dot{x})$ space has been explored for several values of the energy $h$. We managed to distinguish between trapped (non-escaping) and escaping orbits and we also located the basins of escape leading to different exit channels, finding correlations with the corresponding escape times of the orbits. Among the escaping orbits, we separated between those escaping fast or late from the system. Our extensive numerical calculations strongly suggest, that the overall escape process is very dependent on the value of the total orbital energy. We also performed a statistical analysis in each case, relating the proportion of escaping and directly escaping orbits with the value of the energy. In the same vein, the evolution of the proportion of escaping orbits and the corresponding probability, as functions of the $n$th intersection with the $y = 0$ axis upwards was also presented.

The main numerical results of our investigation can be summarized as follows:
\begin{enumerate}
 \item In all three cases studied, areas of trapped orbits and regions of initial conditions leading to escape in a given direction (basins of escape), were found to exist in both the physical and the phase space. The several escape basins are very intricately interwoven and they appear either as well-defined broad regions or thin elongated spiral bands. Regions of trapped orbits first and foremost correspond to stability islands of regular orbits where a third integral of motion is present.
 \item A strong correlation between the extent of the basins of escape and the value of the energy $h$ was found to exists. Indeed, for low values of $h$ the structure of both physical and phase space exhibits a large degree of fractalization and therefore the majority of orbits escape choosing randomly escape channels. As the value of $h$ increases however, the structure becomes less and less fractal and several basins of escape emerge. The extent of these basins of escape is more prominent at high energy levels, where they occupy about 90\% of the entire area on the grids.
 \item It was found, that for energy levels slightly above the escape energy the majority of the escaping orbits have considerable long escape rates (or escape periods), while as we proceed to higher energies the proportion of fast escaping orbits increases significantly. This phenomenon can be justified, if we take into account that with increasing energy the exit channels on the equipotential curves become more and more wide thus the test particles can find easily and faster one of the exits and escape to infinity.
 \item We observed, that in several exit regions the escape process is highly sensitive dependent on the initial conditions, which means that a minor change in the initial conditions of an orbit lead the test particle to escape through another exit channel. These regions are the opposite of the escape basins, are completely intertwined with respect to each other (fractal structure) and are mainly located in the vicinity of stability islands. This sensitivity towards slight changes in the initial conditions in the fractal regions implies, that it is impossible to predict through which exit the particle will escape.
 \item Our calculations revealed, that the escape times of orbits are directly linked to the basins of escape. In particular, inside the basins of escape as well as relatively away from the fractal domains, the shortest escape rates of the orbits had been measured. On the other hand, the longest escape periods correspond to initial conditions of orbits either near the boundaries between the escape basins or in the vicinity of stability islands.
 \item In the case where the perturbation term creates four channels of escape in the physical space, we found that a small portion of chaotic orbits with initial conditions close to the outermost KAM islands remain trapped in the neighbourhood of these islands for vast time intervals having sticky periods which correspond to hundreds of thousands time units. On the contrary in systems with two and three exit channels all non-escaping orbits are regular, while all tested chaotic orbits escape to infinity within the predefined integration time.
\end{enumerate}

We hope that the present review analysis and the corresponding numerical results to be useful in the active field of open Hamiltonian systems which may have implications in different aspects of chaotic scattering with applications in several areas of physics. In Part II of our investigation, we shall try to reveal the escape properties of orbits in dynamical systems with $n$ $(n \geq 5)$ channels of escape in the physical space. Furthermore, it is in our future plans to expand our exploration in other more complicated potentials, focusing our interest in revealing the escape process of orbits of stars in realistic galactic systems (i.e., star clusters, binary stellar systems, rotating galaxies leaking stars, etc).

\section*{Acknowledgments}

The author would like to express his warmest thanks to the two anonymous referees for the careful reading of the manuscript and for all the apt suggestions and comments which allowed us to improve both the quality and the clarity of the paper.

\end{document}